\documentclass[11pt,english,a4paper]{article}

\usepackage[T1]{fontenc}
\usepackage[utf8]{inputenc}

\usepackage{xcolor,soulutf8}

\usepackage{graphicx}
\usepackage{mathtools,amsfonts,amssymb}
\usepackage{mathrsfs} 

\usepackage[margin=10pt,font=small,labelfont=bf,labelsep=endash]{caption} [2023/03/12]

\definecolor{vertfonce}{rgb}{0.20, 0.46, 0.25}
\definecolor{rougefonce}{rgb}{0.64, 0.09, 0.20}

\usepackage{babel}
\usepackage[breaklinks=true,
colorlinks=true,
linkcolor=rougefonce,
citecolor=vertfonce]{hyperref}

\DeclareUnicodeCharacter{1ECD}{\d o}

\newcommand{\RM}{\mathbb{R}}
\newcommand{\dd}[1]{\ensuremath{\operatorname{d}\!{#1}}}
\newcommand{\pscal}[2]{\langle{}#1,\,#2\rangle}
\newcommand{\abs}[1]{\left|#1\right|}
\newcommand{\norm}[1]{\left\|#1\right\|}
\newcommand{\snorm}[1]{\|#1\|}
\newcommand{\ssi}{\Longleftrightarrow}
\newcommand{\tr}{\mathop{\operatorname{tr}}\nolimits}
\newcommand{\restr}{\upharpoonright}
\newcommand{\cqfd}{\hfill $\square$\par\vspace{1ex}}
\newcommand{\Cinf}{C^\infty}
\newcommand{\intint}{\int\!\!\!\int}
\newcommand{\h}{\hbar}
\renewcommand{\O}{\mathcal{O}}
\renewcommand{\geq}{\geqslant}
\renewcommand{\leq}{\leqslant}
\newcommand{\theor}{Theorem}
\newcommand{\defin}{Definition}
\newcommand{\lemma}{Lemma}
\newcommand{\remar}{Remark}
\newcommand{\exemp}{Example}
\newcommand{\corol}{Corollary}
\newcommand{\propo}{Proposition}
\newcommand{\demon}{Proof}
\newtheorem{theo}{\theor}[section]
\newtheorem{theo*}{\theor}

\newtheorem{prop}[theo]{\propo}
\newtheorem{lemm}[theo]{\lemma}
\newtheorem{coro}[theo]{\corol}

\newcommand{\demons}[1][$\!\!$]{\noindent\textbf{\demon\ }\textsl{#1}\textbf{.}~}

\newenvironment{rema}
{\par\vspace{1ex}\refstepcounter{theo}%
\noindent\textbf{\remar~\thetheo} }
{~\hfill\mbox{$\triangle$}\par\vspace{1ex}}

\newenvironment{demo}[1][$\!\!$]
{\demons[#1]\ }
{\cqfd}

\newcommand{\Op}{\mathop{\operatorname{Op}}\nolimits}
 \newcommand{\Tg}{\textup{T}}
\newcommand{\trsp}[1]{{#1}^{\textup{t}}}
\newcommand{\pib}{\boldsymbol{\pi}}

\title{Accurate semiclassical analysis of light propagation on tilted hyperplanes}

\author{Patrick \textsc{Gioia}\footnote{Orange Labs, 2 av. Belle
    Fontaine, F-35510 Cesson-Sévigné.} \and \textsc{Vũ
    Ngọc} 
  San\footnote{Univ Rennes, CNRS, IRMAR --- UMR 6625, F-35000 Rennes,
    France}}

\begin{document}
\maketitle

\begin{abstract}
  In the scalar light model given by Helmholtz' equation in
  $\RM^{1+d}$, we consider the transformation of an initial scene (a
  hologram) in $\{0\}\times \RM^d$ by an arbitrary affine
  transformation (which can be viewed as a propagation into a tilted
  hyperplane).  In the high frequency regime, we use microlocal and
  semiclassical analysis to describe the propagator as a semiclassical
  Fourier integral operator, thus generalising the well-known Angular
  Spectrum formula from optics. We then prove new precise Egorov
  theorems, including subprincipal terms, which indicate how to take
  into account the propagation along rays of geometric optics.
\end{abstract}

\section{Introduction}

The initial motivation for this work has its roots in a technological
challenge currently encountered by researchers in computational
holography. The problem, described for instance
in~\cite{el_rhammad_view-dependent_2018,BLINDER2019}, is to find a new
phase space model for propagating holograms in virtual reality
headsets. This turns out to involve sophisticated mathematical
objects, including semiclassical Fourier integral operators. The goal
of this paper is to present these objects and derive a new formula for
accurately approximating propagated wave functions --- such as
holograms --- in the high-frequency regime. In return, our
mathematical analysis has direct applications to the numerical
treatment of holograms: not only does it provide a new numerical
scheme for coding and propagating holograms
(see~\cite{san-gioia-etc:photonics24, san-gioia-etc:strasbourg24}),
but also it explains artefacts arising in the twisted angular spectrum
method, which stem from the non-injectivity of the canonical
transformation (see Theorem~\ref{theo:canonical}).

\medskip

The mathematical setting is as follows. Let
$\mathcal{U}\subset \RM^{1+d}$, $d\geq 1$, be a domain contained in
the half space $(-C,\infty ) \times \RM^d$ for some constant $C>0$.

For a given $k>0$, let $\psi$ be a non-trivial solution in
$\mathcal{U}$ to the Helmholtz equation
\[
  \Delta\psi + k^2\psi = 0\,.
\]
(Typically, $\psi$ is the monochromatic light wave of frequency
$kc/2\pi$ --- where $c$ is the speed of light in $\mathcal{U}$ ---
emitted from a source object located in the left half-space
$\{(x_0,x_1,\dots, x_d); \ x_0 < -C\}$.)  Our question is the
following: let $\mathcal{G}$ be an affine transformation of
$\RM^{1+d}$; suppose we know only the initial hologram, \emph{i.e.}
the trace of $\psi$ on the hyperplane $P_0:=\{x_0=0\}$ (assuming this
trace is well-defined), then how to describe the trace of
$\psi\circ \mathcal{G}$ on $P_0$ (the transformed hologram)? More
precisely, how to describe the operator
\[
  U_{\mathcal{G}} : \psi_{\restr P_0} \mapsto \psi\circ
  \mathcal{G}_{\restr P_0}\,?
\]
(In the sequel we often call $U_{\mathcal{G}}$ the \emph{propagator}.)

Two special cases are worth mentioning. First, if $\mathcal{G}$ just a
translation along the $x_0$ axis: $\mathcal{G}(X)= X+\gamma$ for some
$\gamma=(\gamma_0,0,\dots, 0)$, then the above question translates
into:
\begin{quote}
  Given a hologram at position $x_0$, how to compute the hologram at
  position $x_0+\gamma_0$?
\end{quote}
\begin{figure}[h]
  \centering
  \includegraphics[width=\linewidth]{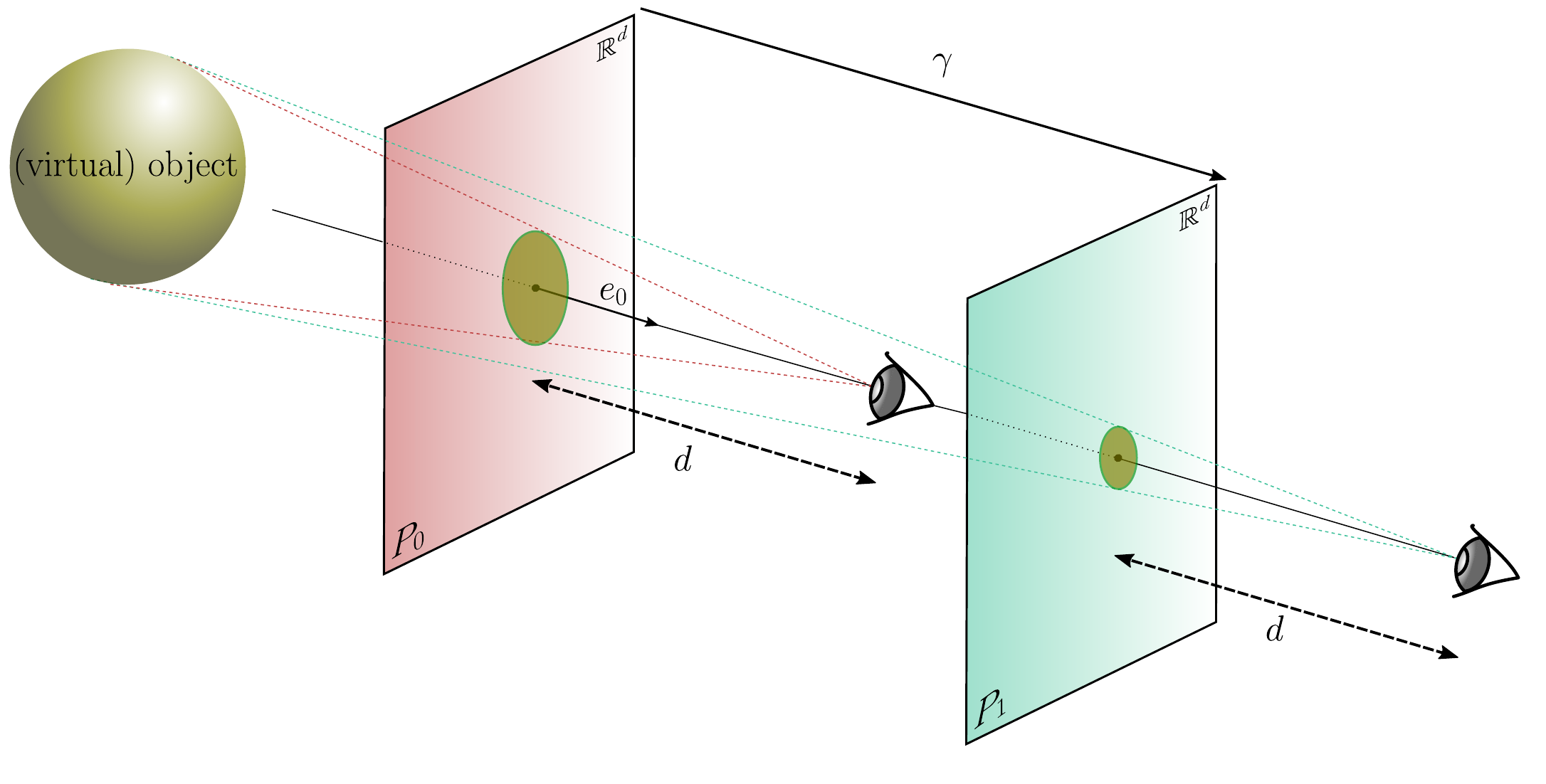}
  \caption{Translating a hologram along the viewer axis. The
    initial hologram is the restriction of the light wave $\psi$
    emitted from the (virtual) object to the plane $P_0$. The
    transformed hologram is the restriction of $\psi\circ \mathcal{G}$
    to $P_0$, which, when $\mathcal{G}(X)=X+\gamma$, is the same as
    the restriction of $\psi$ to the translated plane
    $P_1:=\mathcal{G}(P_0)$. Each hologram is presented to the viewer,
    at a fixed distance $d$ from the eye. When looking at the second
    hologram, the viewer sees the object at a distance increased by
    $\norm{\gamma}$.}\label{fig:translation}
\end{figure}
Thus, a viewer located on the right hand side, looking at the
transformed hologram, will have the impression that the object has
moved backward by a distance $\gamma_0$
(Figure~\ref{fig:translation}). This transformation (a translation
along the viewer axis) is well known in optics, and can be efficiently
computed using the \emph{angular spectrum method} (see
Section~\ref{sec:angular-spectrum}).

The second interesting situation is the case of a Euclidean rotation
$\mathcal{G}\in \textup{SO}(1+d)$.  Of course, an internal rotation
(within the hyperplane $P_0$) is trivial to obtain: it suffices to
rotate the initial hologram. However, when the hyperplane itself is
rotated into a different hyperplane $\mathcal{G}P_0$, then the result
is not obvious. This amounts to answering the question
\begin{quote}
  Given an initial hologram on $P_0$, how to compute the hologram on
  the tilted hyperplane $\mathcal{G}P_0$?
\end{quote}
Thus, a viewer looking at the transformed hologram will have the
impression that the object has \emph{rotated}, or that the object is
fixed while they actually \emph{turn around the object}.

A general affine transformation is a composition of both translations
and rotations, but also of dilations in various directions, which can
emulate the effect of optical lenses. We introduce this ``tilted plane
generalisation'' of the angular spectrum method in
Section~\ref{sec:tilted-planes}.
\begin{figure}[h]
  \centering
  \includegraphics[width=\linewidth]{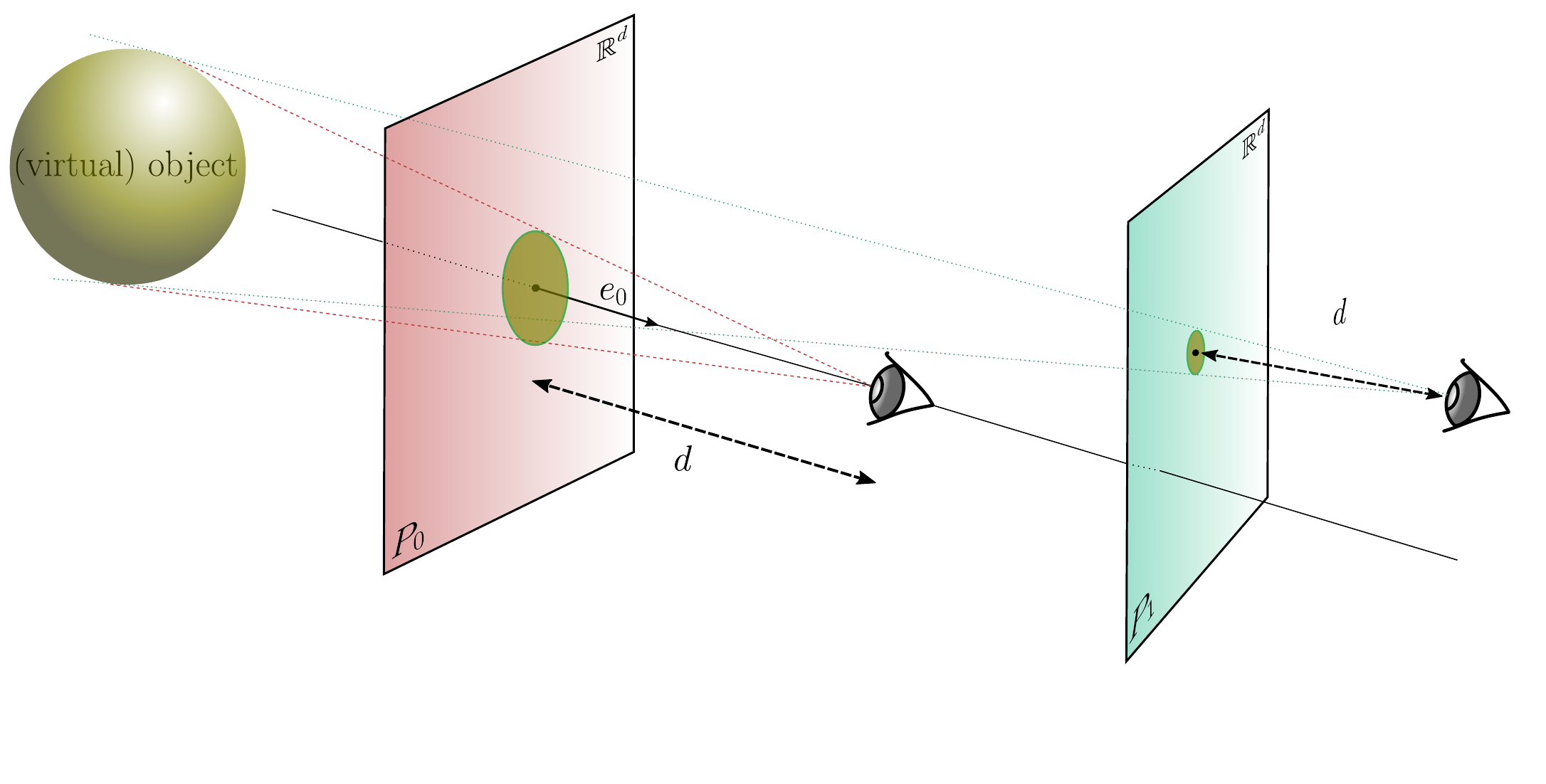}
  \caption{Translating and rotating a hologram. On the
    transformed hologram $P_1$, the viewer now starts to see the
    right-hand side of the object, which was hidden in the initial
    configuration on $P_0$.}\label{fig:rotation}
\end{figure}

\medskip

The goal of this paper is to cast the analysis of the operator
$U_{\mathcal{G}}$ in the framework of \emph{microlocal semiclassical
  analysis}. In the regime of small wavelength (compared to the
details of the object), it is known that a natural and useful approach
to wave optics is to use semiclassical analysis (which was initially
devised for quantum equations with a small ``Planck constant'' $\h$),
see Section~\ref{sec:semiclassical-limit}. The semiclassical analysis
of Helmholtz' equation has been discussed by many authors, see for
instance~\cite{castella05} and references therein. Although we could
not find a precise reference recovering geometric optics from our
Helmholtz propagator, a similar and very nice discussion, based on the
Fresnel formula (which is an approximation to the angular spectrum
formula, in the paraxial regime) can be found
in~\cite{guillemin-sternberg-physics}. Thus, our first result is to
find the conditions under which $U_{\mathcal{G}}$ is a good
semiclassical Fourier integral operator, with a Hörmander
non-degenerate phase function (Section~\ref{sec:four-integr-oper}),
and to determine when it is associated with a canonical transformation
(Proposition~\ref{prop:hessian}); we then compute explicitly its
canonical transformation (Theorem~\ref{theo:canonical}), which in
general is not injective. We explain the geometric meaning of these
results, which allows us to recover the laws of geometric optics in
Section~\ref{sec:geometric-optics}.

The second contribution of this paper, motivated by the initial
hologram analysis challenge, is to obtain precise information on the
transformed hologram $ U_{\mathcal{G}}(\psi_0)$ \emph{without actually
  computing it}. This is achieved first by studying the action of
pseudodifferential operators (viewed as space-frequency filters) on
$U_{\mathcal{G}}$; in Propositions~\ref{prop:KPV} and~\ref{prop:KVQ},
we obtain semiclassical expansions of the Schwartz kernel of the
product of $U_{\mathcal{G}}$ with arbitrary pseudo-differential
operators.  Then, as a consequence, we can predict the observation of
the transformed hologram by means of precise Egorov theorems, keeping
track of all \emph{subprincipal} terms (\emph{i.e.}  terms of order
$\O(\h)$, which correct the first order approximation given by
geometric optics), see Section~\ref{sec:prec-egor-theor}. More
precisely, we prove two Egorov-type formulas with $\O(\h^2)$
remainders: one for $U_{\mathcal{G}}^{-1}P U_{\mathcal{G}}$
(Theorem~\ref{theo:egorov}), and one for
$U_{\mathcal{G}}^* P U_{\mathcal{G}}$
(Theorem~\ref{theo:egorov-adjoint}). Indeed, $U_{\mathcal{G}}$ is not
(even microlocally) unitary --- and the physics literature is often
unclear on this issue. We compute the defect of unitarity in
Section~\ref{sec:lack-unitarity}, which allows us to prove
Theorem~\ref{theo:egorov-adjoint} from Theorem~\ref{theo:egorov}.

\section{Angular spectrum}\label{sec:angular-spectrum}

One of the most used methods for computing the propagation in free
space of a scalar light wave emanating from a 3D scene is called the
Angular Spectrum Method~\cite{goodman-book2005}. It is easy to
implement, well studied, applicable to many situations, and reasonably
fast thanks to the use of the Fast Fourier Transform, see for
instance~\cite{zeng-mcgough09}. In this section, we recall the
well-known Angular Spectrum formula from Fourier Optics, upon which
this method is based. For the purpose of our work, we shall consider a
Euclidean space of arbitrary dimension $1+d$, as this presents no
additional difficulty, although the physically relevant case for
optics is naturally $1+d=3$.

Our initial data $\psi_0(x_1,\dots,x_d)$ will be a screen, or
hologram, which is the trace of a light signal on the hyperplane
$P_0 = \{0\} \times \RM^d$, and our main direction of propagation is
along $x_0$. In the optical community, the coordinates $x_0,x_1,x_2$
of Euclidean $3$-dimensional space are traditionally called $z,x,y$,
and propagation occurs along (or close to) the $z$ axis. We assume
that the light source is located in the left half-space $x_0<0$; as we
shall see, it can be convenient to have a more precise assumption (for
instance, that the light source is compactly supported inside that
half-space).

We work in the scalar wave approximation, and restrict ourselves to
monochromatic waves, with frequency $\omega/2\pi$, \emph{i.e.} of the
form
\[
  (t,x_0,\dots x_d) \mapsto \psi(x_0,\dots,x_d)e^{-i\omega t}\,.
\]
The total wave function $\psi=\psi(x_0,x_1,\dots,x_d)$ is described on
source-free domains $\mathcal{U}\subset \RM^{1+d}$ by the Helmholtz
equation with wave number $k=\omega/c$ ($c$ being the speed of light
in vacuum, and the refractive index is assumed here to be constant
equal to 1), through the incomplete Cauchy problem
\begin{equation}
  \begin{cases}
    \Delta\psi + k^2\psi = 0\\
    \psi_{\restr x_0=0} = \psi_0\,,
  \end{cases}\label{equ:helmholtz}
\end{equation}
where $\Delta=\sum_{j=0}^d\frac{\partial^2}{\partial x_j^2}$ is the
analysts' Laplacian.  In order to perform Fourier analysis, we assume
that $\psi_0\in L^2(\RM^d)$, or that $\psi_0$ is a tempered
distribution. (Of course, since $-\Delta$ has continuous spectrum
$[0,+\infty)$, we cannot expect $\psi$ to be globally in
$L^2(\RM^{1+d})$, if $\psi_0\neq 0$. For instance, it was proven
in~\cite{bonnet-ben-dhia-CRAS16} that, if $\mathcal{U}$ is an open
sector containing a half-space, then the only solution in
$L^2(\mathcal{U})$ is the zero function.) Let $\mathcal{F}_d$ be the
partial Fourier transform in the $(x_1,\dots,x_d)$ variables,
\emph{i.e.}
\[
  (\mathcal{F}_d\psi)(x_0,\zeta_1, \dots, \zeta_d) =
  \int_{\RM^d}e^{-i\pscal{x}{\zeta}_d}\psi(X) \dd x_1 \cdots \dd x_d
  \,,
\]
where
\[
  X := (x_0,x_1,\dots, x_d) \quad x:=(x_1,\dots, x_d) \quad \zeta :=
  (\zeta_1, \dots, \zeta_d)
\]
and
\[
  \pscal{x}{\zeta}_d := \sum_{j=1}^d x_j\zeta_j\,.
\]
In optical terminology, we may refer to $\mathcal{F}_d\psi$ as the
``spectrum'' of $\psi$ on the parallel plane
$P_{x_0} := \{x_0\} \times \RM^d$. By the Fourier inversion formula
applied on each $P_{x_0}$, one has
\begin{equation}
  \psi(x_0,x_1,\dots,x_d) = \frac{1}{(2\pi)^d}\int_{\RM^d} e^{i\pscal{x}{\zeta}_d}
  \mathcal{F}_d\psi(x_0,\zeta_1, \dots, \zeta_d)\dd{\zeta_1} \cdots \dd{\zeta_d}\,.
  \label{equ:ansatz}
\end{equation}
Inserting~\eqref{equ:ansatz} in Helmholtz'
equation~\eqref{equ:helmholtz} and applying $\mathcal{F}_d$ we obtain
\[
  \partial_{x_0}^2\mathcal{F}_d\psi + (k^2 - \norm{\zeta}^2)
  \mathcal{F}_d\psi = 0\,.
\]
In this differential equation, the variables
$(\zeta_1,\dots, \zeta_d)$ can be seen as parameters; this shows that
two qualitatively different regimes may occur depending on the Fourier
domain we are interested in, based on the sign of
$(k^2 - \norm{\zeta}^2)$. If we assume that
\begin{equation}
  \norm{\zeta}^2 \leq k^2
  \label{equ:region}
\end{equation}
then we have an oscillatory solution of the form
\begin{equation}
  \mathcal{F}_d\psi(x_0,\zeta) = A(\zeta) e^{ix_0\sqrt{k^2-\snorm{\zeta}^2}}
  + B(\zeta) e^{-ix_0\sqrt{k^2-\snorm{\zeta}^2}}\,,
  \label{equ:helmholtz-solutions}
\end{equation}
with
$A(\zeta)+B(\zeta) = \hat\psi_0(\zeta) :=
\mathcal{F}_d\psi_0(0,\zeta)$. In order to make an educated guess for
$A$ and $B$, we need to supplement the initial condition with a
physically acceptable condition at infinity. Suppose we have chosen
the square roots in~\eqref{equ:helmholtz-solutions} to lie in the
upper half-plane for negative numbers. Then, in the propagation
direction we are interested in, $x_0>0$, the first exponential
$ e^{ix_0\sqrt{k^2-\snorm{\zeta}^2}}$ gives an evanescent wave when
$k^2<\norm{\zeta}^2$, which is physically acceptable (and
mathematically amenable to the inverse Fourier transform), contrary to
the term $e^{-ix_0\sqrt{k^2-\snorm{\zeta}^2}}$ which grows
exponentially. This invites us to take $A= \hat\psi_0$ and $B=0$,
which yields the so-called Angular Spectrum representation for $\psi$,
see for instance~\cite[3.10.2]{goodman-book2005}:
\begin{equation}
  \psi(x_0,x_1,\dots,x_d) = \frac{1}{(2\pi)^d}\int_{\RM^d}
  e^{i(\pscal{x}{\zeta}_d + x_0\sqrt{k^2-\snorm{\zeta}^2})}
  \hat\psi_0(\zeta)\dd{\zeta}\,,
  \label{equ:angular-spectrum}
\end{equation}
which, as long as the involved quantities stay in a class where the
Fourier transform $\mathcal{F}_d$ is applicable, solves the Helmholtz
equation~\eqref{equ:helmholtz}. In the high frequency regime, global
estimates for Helmholtz' solutions, in more general settings (variable
refraction, limit radiation term) have been studied by many authors,
see~\cite{royer14} and references therein. In fact, as we shall see
below (Remark~\ref{rema:lagrangian}), the choice made in the Angular
Spectrum formula is essentially of microlocal nature, and will have
important implications in our analysis.

\section{Tilted planes}\label{sec:tilted-planes}

\begin{figure}[h]
  \centering
  \includegraphics[width=0.5\textwidth]{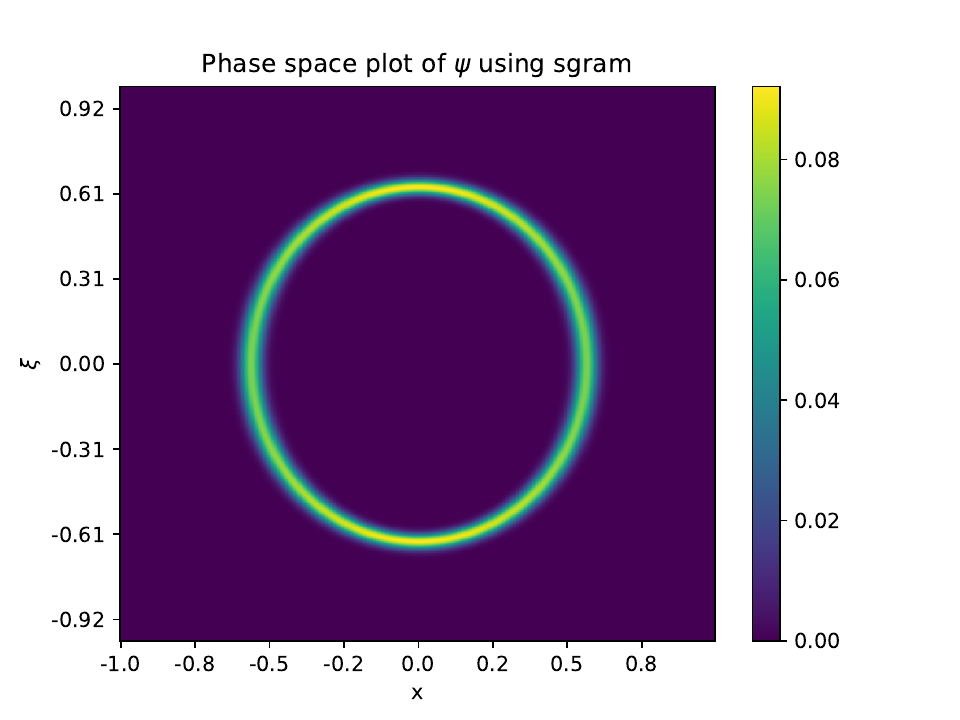}%
  \includegraphics[width=0.5\textwidth]{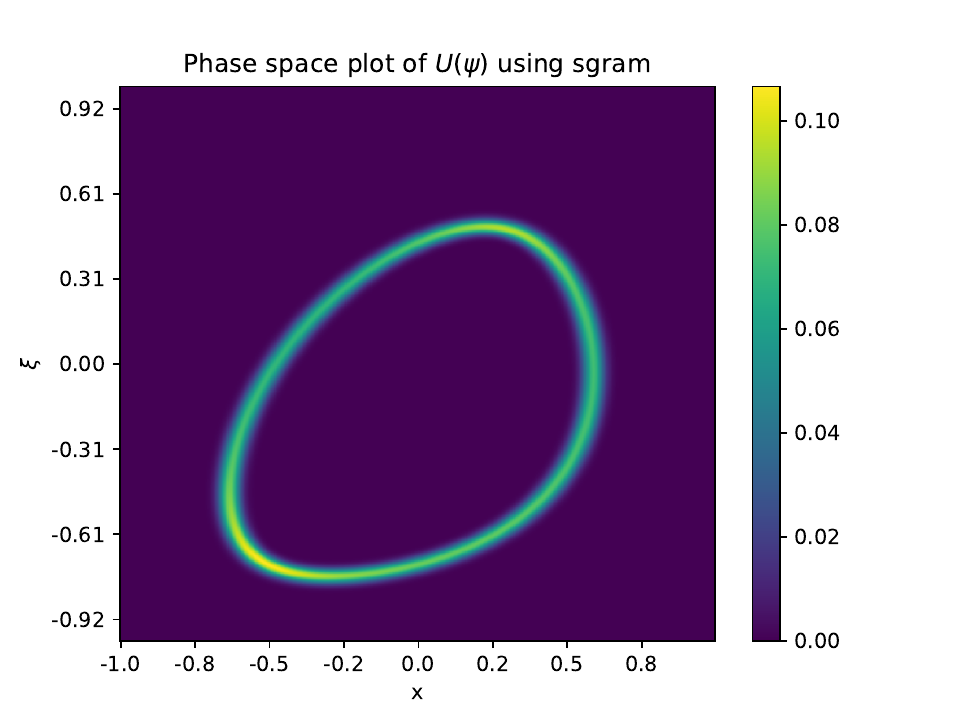}%
  \caption{Here $\psi_0$ is a highly excited 1D Hermite function,
    whose phase space representation is an ellipse (left figure).  On
    the right figure, we display the phase space representation of its
    image $\tilde \psi = \psi\circ \mathcal{G}$, where $\mathcal{G}$
    is a rotation of angle 10° followed by a translation of vector
    $\gamma=(0.3, 0)\in\RM^2$, obtained by implementing
    Formula~\eqref{equ:tilted-angular-spectrum}. The shape of the
    transformed signal will be explained by
    Theorem~\ref{theo:canonical}, see
    Figure~\ref{fig:hermite-egorov}. The phase space representations
    are obtained using the \texttt{sgram} command from the
    \texttt{ltfat} library.}\label{fig:hermite}
\end{figure}

In this work, we are interested in the restriction of the light signal
$\psi$ from~\eqref{equ:angular-spectrum} to a (tilted) hyperplane, and
more precisely in the map that sends $\psi_0$ to that restriction, see
Figure~\ref{fig:hermite}.

Let $(e_0,e_1,\dots, e_d)$ be the canonical basis of $\RM^{1+d}$, and
denote the corresponding coordinates by $ X=(x_0,x_1,\dots, x_d)$. A
general hyperplane $P_{a,\beta}\subset \RM^{1+d}$ is defined by the
equation
\begin{equation}
  \pscal{a}{X} = \beta\label{equ:hyperplane}\,,
\end{equation}
where $a\in S^d$ is a unit vector in $\RM^{1+d}$ and $\beta \geq 0$.
Let us pick up an orthogonal transformation $G\in \textup{SO}(d+1)$
such that $a = {\trsp G}^{-1} e_0 = Ge_0$. Recall that our reference
hyperplane is $P_0 = \{0\} \times \RM^d$. Since $P_0 = P_{e_0,0}$, it
follows from~\eqref{equ:hyperplane} that
\[
  P_{a,\beta} = G(P_0 + \beta e_0)\,.
\]
Let $\mathcal{G}:= \tilde X\mapsto G(\tilde X+\beta e_0)$. The affine
map $\mathcal{G}:\RM^{1+d}\to\RM^{1+d}$ is a diffeomorphism that sends
$P_0$ to $P_{a,\beta}$, so we may use it to parameterise $P_{a,\beta}$
by $P_0$ through the new coordinates
$\tilde X=(\tilde x_0,\tilde x_1,\dots,\tilde x_d)$ defined by
\[
  X = \mathcal{G} \tilde X\,.
\]

Consider now Equation~\eqref{equ:angular-spectrum}, which we may
rewrite as
\[
  \psi(X) = \frac{1}{(2\pi)^d}\int_{\RM^d}e^{i\pscal{X}{Z}}
  \hat\psi_0(\zeta) \dd\zeta, \qquad X\in \mathcal{H}_+,
\]
where $\mathcal{H}_+$ is the right half-space:
$\mathcal{H}_+ = \{X\in\RM^{1+d}, x_0\geq 0\}$, and
\begin{equation}
  Z = Z(\zeta) = \left(\sqrt{k^2-\snorm{\zeta}^2}, \zeta_1, \dots,
    \zeta_d\right)\,.\label{equ:sigma_first}
\end{equation}
Expressing $\psi$ in the new coordinates (\emph{via}
$\tilde\psi(\tilde X) = \psi(X)$) we get
$\tilde\psi = \psi\circ \mathcal{G}$.  Hence the expression of the
angular spectrum restricted to the new hyperplane $P_{a,\beta}$, in
the new coordinates $\tilde X = (0,\tilde x_1, \dots, \tilde x_d)$, is
simply
\begin{equation}
  \tilde \psi(\tilde X) = \psi(\mathcal{G}(\tilde X)) =
  \frac{1}{(2\pi)^d}\int_{\RM^d}e^{i\pscal{\mathcal{G}(\tilde{X})}{Z}}
  \hat\psi_0(\zeta)
  \dd\zeta\,.
  \label{equ:tilted-angular-spectrum}
\end{equation}
We shall call this formula the ``rotated angular spectrum''. It holds
in general only when $\mathcal{G}(\tilde X)\in \mathcal{H}_+$; if we
assume
$\mathop{\textup{supp}}\hat\psi_0\subset \{\norm{\zeta}^2\leq k^2\}$
(see Section~\ref{sec:semiclassical-limit} for a justification of this
hypothesis), it can be extended to any $\tilde X\in \RM^{1+d}$,
however this is physically relevant only when $\mathcal{G}(\tilde X)$
stays in a source-free region.

\begin{rema}\label{rema:jacobian}
  In the 3D case ($d=2$), a similar analysis was carried out
  in~\cite{matsushimaFastCalculationMethod2003} for tilted planes with
  no translation ($\beta=0$). However, an incorrect formula (similar
  to~\eqref{equ:tilted-angular-spectrum}, but involving a superfluous
  Jacobian) was presented, and this was partly clarified in a
  subsequent paper~\cite{matsushima_formulation_2008}: since
  $\beta=0$, one can conveniently express $\tilde\psi$ as the inverse
  Fourier transform of some planar signal.  This alternative formula,
  which we extend here to any dimension, requires an additional
  geometric assumption on $P_{a,\beta}$; indeed, since
  $\mathcal{G}=G$, we can write
  \[
    \tilde \psi(\tilde X) =
    \frac{1}{(2\pi)^d}\int_{\RM^d}e^{i\pscal{\tilde{X}}{\trsp G Z}}
    \hat\psi_0(\zeta) \dd\zeta\,,
  \]
  where $\trsp G$ is the adjoint of $G$ for the Euclidean structure.
  Since $\tilde X = (0,\tilde x)$, we see that the scalar product in
  the phase involves only the last $d$ components of $\trsp G Z$.
  Now, assume that $\trsp G Z$ induces a diffeomorphism on the
  hyperplane $\{0\}\times\RM^d$, \emph{i.e.}  there is a
  diffeomorphism $g:\RM^d\to\RM^d$ such that
  $\trsp G Z(\zeta) = (\tilde t(\zeta), g(\zeta))$ for some smooth
  function $\tilde t$. Under this assumption, one could make the
  change of variables $\tilde \zeta = g(\zeta)$:
  \[
    \begin{aligned}
      \tilde \psi(\tilde X)
      & =
        \frac{1}{(2\pi)^d}\int_{\RM^d}e^{i\pscal{\tilde{X}}{\tilde Z}}
        \hat\psi_0(\zeta)
        \dd\zeta\ \quad \text{ with } \tilde Z = \trsp G Z\\
      & = \frac{1}{(2\pi)^d}\int_{\RM^d} e^{i\pscal{\tilde x}{\tilde{\zeta}}_d}
        \hat\psi_0(g^{-1}(\tilde\zeta)) J(\tilde{\zeta})\dd{\tilde{\zeta}} \\
      & = \mathcal{F}_d^{-1}((\hat\psi_0\circ g^{-1}) J)(\tilde x)\,,
    \end{aligned}
  \]
  where $J(\tilde\zeta)$ is the Jacobian of $g^{-1}$. In other words,
  viewing $J$ as a multiplication operator, we would have
  \begin{equation}
    G^*\psi(0,\tilde x) = \mathcal{F}_d^{-1}\circ J \circ (g^{-1})^*
    \circ  \mathcal{F}_d \psi_0 (\tilde x).
    \label{equ:matsushima-factorized}
  \end{equation}
  (We use the superscript ``$\,^*\,$'' for the pull-back operator, for
  instance:
  \[ {G}^{*}\psi = \psi\circ{G}\,).
  \]

  However, this formula is still not correct, simply because the
  assumption that $g$ is a diffeomorphism can never be realised, due
  to the square root in~\eqref{equ:sigma_first}, see
  Section~\ref{sec:study-v_G}, which brings several difficulties in:
  first, one has to assume that $\hat \psi_0$ should be supported in
  the unit disc, otherwise $Z$ might take complex values; secondly,
  and most importantly, $g$ is generally not injective, but rather ``2
  to 1''. Hence one has to carefully choose an integration sheet in
  the spectral variable. For these reasons, we see that a
  ``microlocal'' (\emph{i.e.} local in phase space) approach will be
  more appropriate for a good understanding of this transform. On the
  numerical side, failure to take into account the non-injectivity may
  result in parts of the viewed objects being hidden as one performs a
  rotation via the angular spectrum method.

  Notice also that, if $g$ were a linear transformation, the
  right-hand side of~\eqref{equ:matsushima-factorized} would simply be
  equal to $g^* \psi_0 (\tilde x)$. But, of course, even when $G$ is
  linear, $g$ itself is generally non-linear, due to the square root
  in~\eqref{equ:sigma_first}. In some sense, it is the goal of this
  paper, in addition to the introduction of the microlocal setting, to
  deal with this non-linearity.
\end{rema}

\begin{rema}
  Since, in this section, we chose $G$ to be an orthogonal matrix, we
  could replace $\trsp G$ by $G^{-1}$; however, keeping the notation
  $\trsp G$ will allow us to consider general linear transformations,
  which corresponds to additional scaling and shearing of the
  hologram; this notation also reminds us that, while $G$ acts on
  position variables $(x_0,\dots, x_d)$, the transpose $\trsp G$
  should act on frequency (\emph{i.e.} ``Fourier'') variables
  $(\xi_0,\dots, \xi_d)$.  Note however that we use the same canonical
  basis $(e_0,\dots, e_d)$ for both position and Fourier spaces
  $\RM^{1+d}$.
\end{rema}
\begin{rema}
  The map $\mathcal{G}$ preserves $P_0$ if and only if $\beta=0$ and
  the linear part $G$ preserves $P_0$ (and hence $Ge_0 = \pm
  e_0$). This is exactly when the map $g$ of
  Remark~\ref{rema:jacobian} is linear. Hence, in this case,
  \[
    \tilde\psi(0,\tilde x) = \psi_0(g (\tilde x))
  \]
  \emph{i.e.}  we simply perform a rotation within the initial state
  $\psi_0$.

  Similarly, the hyperplane $P_{a,\beta}$ is \emph{parallel} to $P_0$
  if and only if the linear part $G$ preserves $P_0$ (and hence
  $Ge_0 = \pm e_0$). In this case again, $G$ acts on $P_0$ as a linear
  map $g$, and Formula~\eqref{equ:tilted-angular-spectrum} becomes
  \[
    \tilde \psi(0,\tilde x) = \mathcal{F}_d^{-1} \left(
      e^{i\beta \sqrt{k^2-\snorm{\zeta}^2}} \mathcal{F}_d
      \psi_0\right)(g(\tilde x))\,,
  \]
  which, modulo the internal rotation $g$, is the original Angular
  Spectrum formula~\eqref{equ:angular-spectrum}.
\end{rema}
\begin{rema}
  It would be interesting to treat the case of a curved deformation of
  the initial hyperplane $P_0$, \emph{i.e.} when the transformation
  $\mathcal{G}$ is non-linear.
\end{rema}

\section{Semiclassical limit}\label{sec:semiclassical-limit}

Having in mind the natural regime when the spacial frequency $k$ is
very large, we write $k=\frac{1}{\h}$ for a small parameter $\h>0$ (in
physical terms, the ``Planck constant'' $\h$ used here is effectively
the wave length $\lambda$). Helmholtz' equation becomes
\begin{equation}
  \h^2\Delta\psi + \psi = 0\,.\label{equ:schrodinger}
\end{equation}
The analogy between the semiclassical limit in quantum mechanics and
the limit of geometric optics from Fourier optics has been known for a
long time, see for
instance~\cite{guillemin-sternberg-physics,testorfPhasespaceOptics2010}.
However, it seems that the geometric content of the (rotated) Angular
Spectrum formulas that we develop here, based on the semiclassical
intuition, was not investigated before.

\begin{figure}
  \centering \includegraphics[width=0.8\linewidth]{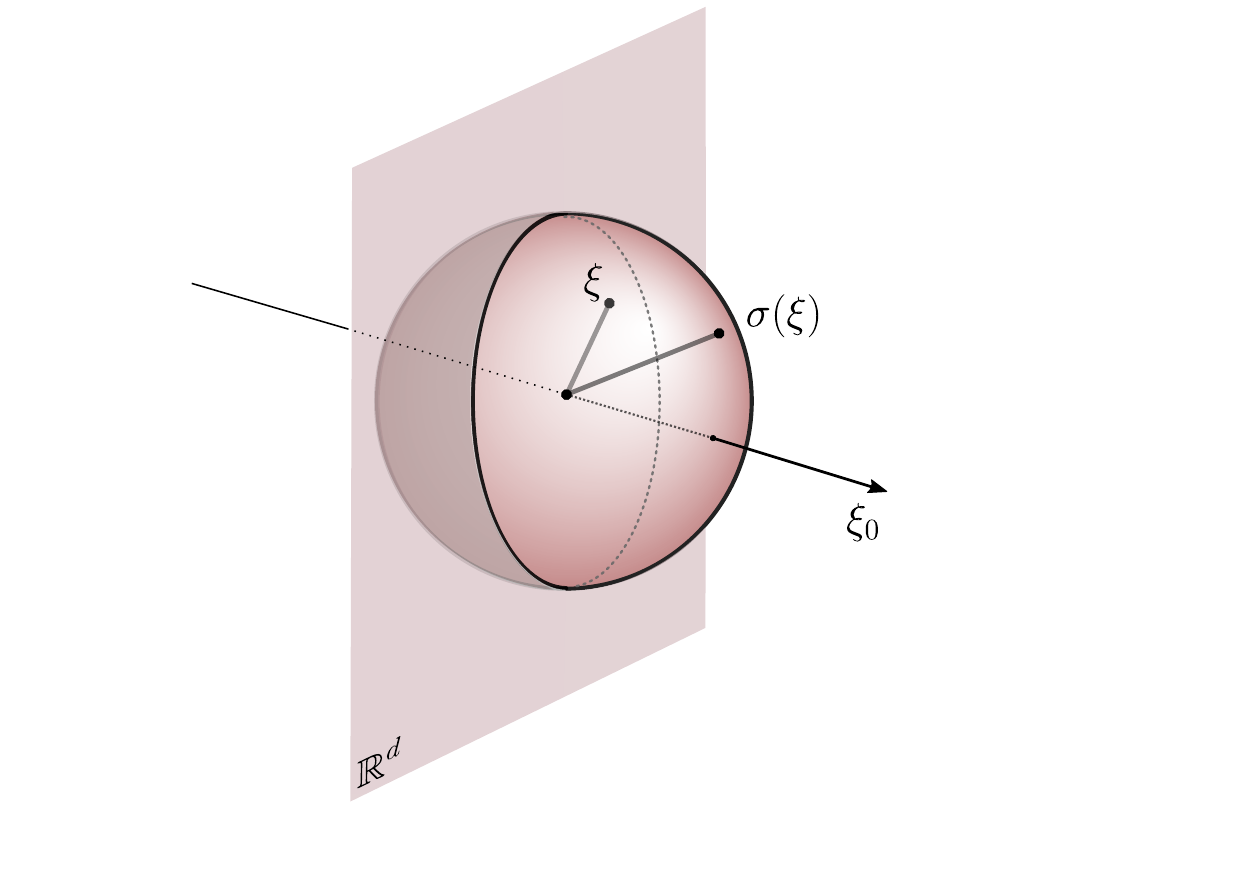}
  \caption{The map $\sigma$}\label{fig:sigma}
\end{figure}

To start with, since Equation~\eqref{equ:schrodinger}, \emph{i.e.}
$-\h^2\Delta \psi = \psi$, is nothing but a semiclassical Schrödinger
equation at energy $1$, this suggests that, in the limit $\h\to 0$,
solutions have to be microlocalized on the unit sphere in the
classical momentum variables $(\xi_0,\dots, \xi_d)$. To be precise,
let us introduce the corresponding scaling, that is,
$\xi = (\xi_1, \dots, \xi_d) = \h\zeta$. Since the semiclassical wave
front of $\psi$ is contained in the sphere $\sum_{j=0}^d \xi_j^2=1$,
we must have $\abs{\xi_0}<1$ in the physical region. Then
\[
  Z_0 = \sqrt{k^2-\snorm{\zeta}^2} = \h^{-1}\sqrt{1 -
    \snorm{\xi}^2}\,,
\]
and the unit sphere in the full semiclassical Fourier variables
$(\h Z_0,\xi_1,\dots, \xi_d)$ corresponds to the unit disc
$\norm{\xi}^2\leq 1$, which is the oscillatory
region~\eqref{equ:region} of Helmholtz' equation, and the ``physically
accessible'' region of phase space for the classical mechanical limit
of Schrödinger's equation at energy 1. Up to an $\O(\h^\infty)$ error,
we may truncate $\psi$ in the neighbourhood of this unit co-sphere,
which ensures that the trace $\psi_0$ on the vertical hyperplane
$\mathcal{P}_0:=\{\xi_0=0\}$ is well-defined. The rotated Angular
Spectrum formula~\eqref{equ:tilted-angular-spectrum} can be written as
\[
  \tilde \psi(\tilde X) = \frac{1}{(2\pi\h)^d}\int_{\RM^d}
  e^{\frac{i}{\h}\pscal{\mathcal{G}(\tilde{X})}{\sigma(\xi)}} \hat\psi_0(\xi/\h)
  \dd\xi\,,
\]
where $\tilde X = (0,\tilde x_1,\dots,\tilde x_d)\in P_0$ and
\begin{equation}
  \sigma(\xi) = \left(\sqrt{1 - \snorm{\xi}^2}, \xi\right)\in S^d \subset \RM^{1+d}\,.
  \label{equ:phy}
\end{equation}
The map $\sigma$ will play an important role in our analysis; it lifts
a vector $\xi\in\RM^d$ from $\mathcal{P}_0$ to the ``right
hemisphere'' ($\xi_0\geq 0$) of the unit sphere $S^d$ in $\RM^{1+d}$,
see Figure~\ref{fig:sigma}.

Notice also that $\hat\psi_0(\xi/\h) = \mathcal{F}_\h\psi_0(\xi)$,
where $\mathcal{F}_\h$ is the semiclassical Fourier transform given by
\[
  \mathcal{F}_\h = \int_{\RM^d}e^{-\frac{i}{\h} \pscal{x}{\xi}} \dd x
\]
(whose inverse is
$\mathcal{F}_\h^{-1} = \frac{1}{(2\pi\h)^d}\mathcal{F}_\h^*$, where
$\mathcal{F}_\h^*$ is the $L^2$-adjoint).  This yields the
semiclassical expression for the value of the wave function on the
tilted hyperplane:
\begin{align}
  \tilde \psi(0,\tilde x)
  & =
    \frac{1}{(2\pi\h)^d}\int_{\RM^d}
    e^{\frac{i}{\h}\pscal{\mathcal{G}\tilde{X}}{\sigma(\xi)}}
    \mathcal{F}_\h\psi_0(\xi) \dd\xi \label{equ:semiclassical-angular1}\\
  & = \frac{1}{(2\pi\h)^d}\int_{\RM^d\times\RM^d}
    e^{\frac{i}{\h}(\pscal{\mathcal{G}\tilde{X}}{\sigma(\xi)} - \pscal{x}{\xi}_d)}
    \psi_0(x) \dd x \dd{\xi}\,.\label{equ:semiclassical-angular2}
\end{align}
As we shall now investigate, from this formula, the operator that
sends $\psi_0$ to the map $\tilde x \mapsto \tilde \psi(0,\tilde x)$
has the structure of a semiclassical Fourier integral operator. As a
byproduct, we will recover, without imposing any axiom, the usual
structures of geometric optics, thanks to the symplectic geometry
obtained in the limit $\h\to 0$.

\section{Fourier integral operators}\label{sec:four-integr-oper}

Fourier integral operators, or FIOs, are (microlocal) operators $U$
acting on functions $\psi$, of the form
\begin{equation}
  (U\psi)(\tilde x) = \intint e^{\frac{i}{\h}\phi(\tilde x,x,\theta)}
  a_\h(\tilde x,x,\theta)\psi(x) \dd{x} \dd{\theta}\,,
  \label{equ:oif}
\end{equation}
where $(\tilde x,x,\theta)\in\RM^d \times \RM^d \times \RM^N$ (the
``auxiliary dimension'' $N\leq d$ may depend on the operator), $a_\h$
is called the amplitude and $\phi$, the phase. A relatively recent
account of FIOs can be found for
instance~\cite{guillemin-sternberg-semiclassical}, but most of the
theory was initially developed by Maslov~\cite{maslov72} and
Hörmander~\cite{FIO1} (for the latter, without the small parameter
$\h$); see also~\cite{duistermaat-oscillatory,duistermaat-oif}, and
the nice introductory paper~\cite{guillemin-25years-FIO}. Since the
beginning of microlocal analysis, FIOs have been used to simplify many
partial differential equations by ways of \emph{normal forms},
see~\cite{san-panoramas} and references therein. Interestingly, more
recently, FIOs have also become part of the signal processing toolbox
and can be analysed (theoretically and numerically) via Gabor frames,
see for example~\cite{cordero-2013}, and specifically for the
Helmholtz equation in~\cite{chaumont-dolean-ingremeau24}.

In order to have a good FIO calculus, one asks that the phase be
non-degenerate (in the sense of Hörmander), which essentially means
that the set
\[
  C_\phi := \{(\tilde x,x,\theta) \in \RM^d \times \RM^d \times \RM^N;
  \quad \partial_\theta\phi(\tilde x,x,\theta) = 0 \}
\]
must be a smooth manifold of dimension $2d$. (In view of the
stationary phase lemma, we see that the Schwartz kernel of the
integral operator~\eqref{equ:oif} can be reduced, modulo a very small
error of order $\O(\h^\infty)$, to a ``usual'' kernel depending only
on the variables $(\tilde x, x)$. In other words, the Hörmander
condition allows us to get rid of the auxiliary variable $\theta$.)
What is important for us is that with such an operator is associated a
\emph{canonical relation} in phase space, given by the image
$\Lambda_\phi\subset \RM^{2d} \times \RM^{2d}$ of the map
\[
  C_\phi \ni (\tilde x,x,\theta) \mapsto (\tilde x,\partial_{\tilde x}
  \phi,\, x, -\partial_x\phi)\,.
\]

The canonical relation $\Lambda_\phi$ is always Lagrangian in
$\RM^{2d} \times \RM^{2d}$ (with the symplectic form
$\dd{\tilde{\xi}}\wedge \dd{\tilde x} - \dd{\xi}\wedge \dd x$), but is
not necessarily the graph of an honest transformation. For this to
hold locally, it is sufficient that the matrix
$\partial_{\tilde x}\partial_x \phi$ be everywhere invertible. Since
in our case $x$ and $\tilde x$ have the same dimension, it will then
imply that the canonical transformation is locally invertible: it is a
local symplectic diffeomorphism
$\kappa_\phi: (x, -\partial_x\phi) \mapsto (\tilde x,\partial_{\tilde x}
\phi)$.

The best known example of a non-trivial Fourier Integral Operator is
the semiclassical Fourier transform $\mathcal{F}_\h$: there,
$\phi(\tilde x, x) = -\pscal{\tilde x}x$ (there is no auxiliary
variable $\theta$, hence it is automatically Hörmander
non-degenerate), and $\partial_{\tilde x}\partial_x \phi$ is minus the
identity matrix. We have
\[
  \Lambda_\phi = (\tilde x, -x,\, x, \tilde x)
\]
so the associated symplectic transformation is
$(x,\tilde x)\mapsto (\tilde x, -x)$.

In both formulas~\eqref{equ:semiclassical-angular1}
and~\eqref{equ:semiclassical-angular2}, we have obtained the target
signal $\tilde \psi(\tilde x_1,\tilde x_2, 0)$ from the original one
$\psi_0(x_1,x_2)$ by applying integral operators, which have the form
of Fourier integral operators. Both formulas are actually interesting,
and the goal of this section is to study them in details. While we
believe that these results are new, we note that the use of FIOs for
wave equations has a long tradition, and related formulas can be found
in recent works like~\cite{hillairet-wunsh20}. Moreover, in the
paraxial regime ($\psi_0$ is nearly co-normal), the Helmholtz equation
becomes a time-dependent Schrödinger equation
(see~\cite{doumic-2003,doumic-2009}), whose propagator has naturally
the structure of a semiclassical Fourier integral operator, see for
instance~\cite{robert}.

\begin{rema}\label{rema:lagrangian}
  By definition, the Schwartz kernel $K_U$ of $U$ is called a
  Lagrangian distribution of $(\tilde x,x)$:
  \[
    K_U(\tilde x, x) = \intint e^{\frac{i}{\h}\phi(\tilde x,x,\theta)}
    a_\h(\tilde x,x,\theta)\dd{\theta}\,,
  \]
  associated with the Lagrangian submanifold $\Lambda_\phi$.  Let us
  compare to the choice we made for the Helmholtz
  solution~\eqref{equ:angular-spectrum}, which writes
  \[
    \psi(x_0,x_1,\dots,x_d) = \frac{1}{(2\pi\h)^d}\int_{\RM^d}
    e^{\frac{i}{\h}(\pscal{x}{\xi}_d + x_0\sqrt{1-\snorm{\xi}^2})}
    \hat\psi_0(\xi/\h)\dd{\xi}\,.
  \]
  If we assume that $\hat\psi_0(\xi/\h)=\chi(\xi)$ for some smooth
  function $\chi$, then we see that $\psi$ is a Lagrangian
  distribution of $x$ (also called Maslov-WKB state) whose Lagrangian
  submanifold is contained in the set
  $\{((x_0,x),(\xi_0,\xi))\in T^*\RM^{1+d}, \ \xi_0 =
  \sqrt{1-\snorm{\xi}^2}\}$: the frequency variable $(\xi_0,\xi)$
  belongs to the \emph{right} hemisphere $\{\norm{\xi}=1,\
  \xi_0>0\}$. This particular choice of Lagrangian submanifold of the
  unit cosphere bundle underpins our entire analysis; it correspond to
  propagation in the $x_0>0$ direction (see
  Section~\ref{sec:geometric-optics}), and induces a kind of
  \emph{caustic} behaviour (in the frequency variable) at
  $\norm{\xi}=1$.
\end{rema}

\subsection{Decomposition of the Angular Spectrum propagation}

We now consider a general affine transformation $\mathcal{G}$ of
$\RM^{1+d}$, of the form $\mathcal{G}(\tilde X)= G\tilde X + \gamma$,
where $G\in\textup{GL}_{1+d}(\RM)$ is not necessary orthogonal. This
allows us to shrink or expand the light rays in specific directions,
in addition to the rotation component.

Let us denote by $U_{\mathcal{G}}$ the operator given
by~\eqref{equ:semiclassical-angular2}, that is:
\[
  [U_{\mathcal{G}}\psi](\tilde x) =
  \frac{1}{(2\pi\h)^d}\int_{\RM^d\times\RM^d}
  e^{\frac{i}{\h}(\pscal{\mathcal{G}\tilde{X}}{\sigma(\xi)} -
    \pscal{x}{\xi}_d)} \psi(x) \dd x \dd{\xi}\,.\
\]
It can be interesting to express $U_{\mathcal{G}}$ as a composition of
simpler operators. For instance, it is clear
from~\eqref{equ:semiclassical-angular1} that
$U_{\mathcal{G}} = V_{\mathcal{G}} \circ \mathcal{F}_\h$, with
\begin{equation} [V_{\mathcal{G}} \phi](\tilde x) =
  \frac{1}{(2\pi\h)^d}\int_{\RM^d}
  e^{\frac{i}{\h}\pscal{\mathcal{G}\tilde{X}}{\sigma(\xi)}} \phi(\xi)
  \dd \xi
  \label{equ:VG}
\end{equation}

Remark that if $\mathcal{G}=\textup{Id}$, then
$V_{\mathcal{G}} = \mathcal{F}_\h^{-1}$, so
$U_{\mathcal{G}}=\textup{Id}$.
\begin{lemm}\label{lemm:decomposition}
  For $G\in\textup{GL}_{1+d}(\RM)$, we identify $G$ with the
  transformation $\tilde X \mapsto G\tilde X$, and for
  $\gamma\in\RM^{1+d}$, we let
  $\tau_\gamma:\tilde X \mapsto \tilde X + \gamma$. To simplify
  notation, we write $U_\gamma$ for $U_{\tau_\gamma}$. The following
  holds formally:
  \begin{enumerate}
  \item\label{item:gammas} For any $\gamma_1,\gamma_2\in \RM^{1+d}$,
    $U_{\gamma_1} \circ U_{\gamma_2} = U_{\gamma_2+\gamma_1} $.
  \item\label{item:gamma-G} For any $G\in\textup{GL}_{1+d}(\RM)$ and
    $\gamma\in\RM^{1+d}$,
    \[
      U_{\tau_{\gamma}\circ G} = V_G
      e^{\frac{i}{\h}\pscal{\gamma}{\sigma(\cdot)}} \mathcal{F}_\h =
      U_G U_{\gamma}
    \]
  \item\label{item:G-gamma} If $G\in\textup{GL}_{1+d}(\RM)$ and
    $\gamma\in\RM^{1+d}$, then
    $U_{G\circ \tau_{\gamma}} = U_G U_{G\gamma}$ is in general
    different from $U_{\gamma} U_G$. However, let
    $\mathcal{H}_\pm = \{X\in\RM^{1+d}, \pm x_0\geq 0\}$. If
    ${\trsp G} \sigma (\textup{supp}\,\mathcal{F}_\h\psi)\subset
    \mathcal{H}_+$, then
    \[
      U_{G\circ \tau_{\gamma}} \psi = U_{\gamma} U_G \psi\,;
    \]
    and if
    ${\trsp G} \sigma (\textup{supp}\,\mathcal{F}_\h\psi)\subset
    \mathcal{H}_-$, then
    \[
      U_{G\circ \tau_{\gamma}} \psi = U_{S_0(\gamma)} U_G \psi\,,
    \]
    where $S_0(\gamma)$ is the symmetric to $\gamma$ with respect to
    $\mathcal{P}_0$, \emph{i.e.}
    $S_0(\gamma) = (-\gamma_0,\gamma_1,\dots,\gamma_d)$.
  \end{enumerate}
\end{lemm}
\begin{demo}
  Point~\ref{item:gammas} follows from the formula
  \begin{equation}
    U_\gamma = \mathcal{F}_\h^{-1}
    e^{\frac{i}{\h}\pscal{\gamma}{\sigma(\cdot)}} \mathcal{F}_\h\,.
    \label{eq:U-gamma}
  \end{equation}
  We turn to point~\ref{item:gamma-G}.  Let
  $\mathcal{G}=\tau_\gamma \circ G = \tilde X\mapsto G\tilde X +
  \gamma$. We have $U_G = V_G \circ \mathcal{F}_\h$ and, since
  $\tilde X = (0,\tilde x)$, we get
  $\pscal{\tilde{X}}{\sigma(\xi)} = \pscal{\tilde x}{\xi}_d$, which
  gives
  \begin{equation}
    V_\gamma = \frac{1}{(2\pi\h)^d}\int_{\RM^d}
    e^{\frac{i}{\h}(\pscal{\tilde{X}}{\sigma(\xi)} +
      \pscal{\gamma}{\sigma(\xi)})} \dd \xi = \mathcal{F}_\h^{-1}
    \circ e^{\frac{i}{\h}\pscal{\gamma}{\sigma(\cdot)}}\label{equ:V-gamma}
  \end{equation}
  where $e^{\frac{i}{\h}\pscal{\gamma}{\sigma(\cdot)}}$ is viewed as a
  multiplication operator. From~\eqref{eq:U-gamma}, we have
  \[
    \begin{aligned}
      U_G U_\gamma & = V_G \mathcal{F}_\h \mathcal{F}_\h^{-1}
                     e^{\frac{i}{\h}\pscal{\gamma}{\sigma(\cdot)}} \mathcal{F}_\h \\
                   & = V_G e^{\frac{i}{\h}\pscal{\gamma}{\sigma(\cdot)}}
                     \mathcal{F}_\h\,,
    \end{aligned},
  \]
  and we notice that
  \[
    \begin{aligned}
      V_G (e^{\frac{i}{\h}\pscal{\gamma}{\sigma(\cdot)}} \phi)
      & =
        \frac{1}{(2\pi\h)^d}\int_{\RM^d} e^{\frac{i}{\h}(\pscal{G
        \tilde{X}}{\sigma(\xi)} +
        \pscal{\gamma}{\sigma(\xi)})} \phi(\xi) \dd \xi\\
      & = \frac{1}{(2\pi\h)^d}\int_{\RM^d}
        e^{\frac{i}{\h}\pscal{\mathcal{G}\tilde{X}}{\sigma(\xi)}}
        \phi(\xi)
        \dd \xi \\
      & = V_{\mathcal{G}} \phi\,,
    \end{aligned}
  \]
  so
  $ U_G U_\gamma = V_{\mathcal{G}} \mathcal{F}_\h = U_{\mathcal{G}}$.

  Let us now consider Point~\ref{item:G-gamma}, \emph{i.e.} the
  reverse composition $U_\gamma U_G$. We have
  \[
    [U_\gamma U_G \psi] (\tilde x) = \mathcal{F}_\h^{-1}
    e^{\frac{i}{\h}\pscal{\gamma}{\sigma(\cdot)}}\mathcal{F}_\h\left[
      \tilde y \mapsto \int e^{\frac{i}{\h}\pscal{G\tilde Y}{\sigma(\eta)}}
      \mathcal{F}_\h \psi (\eta)
      \dd{\eta}\right]\,,
  \]
  where $\tilde Y := (0,\tilde y)$.  Let us introduce the linear
  projection from $\RM^{1+d}$ onto the vertical hyperplane
  $\mathcal{P}_0=\{\xi_0=0\}$.
  \[
    \pib : \RM^{1+d} \to \RM^d\,,
  \]
  so that $\forall \xi\in\RM^d, \,\pib(\sigma(\xi)) = \xi$.  We have
  $\pscal{G\tilde Y}{\sigma(\eta)} = \pscal{\tilde y}{\pib {\trsp
      G}\sigma(\eta)}_d$, and hence
  \begin{equation} [U_\gamma U_G \psi] (\tilde x) = \int %
    e^{\frac{i}{\h}\pscal{\tilde x}{\xi}_d} %
    e^{\frac{i}{\h}\pscal{\gamma}{\sigma(\xi)}} %
    e^{\frac{i}{\h} \pscal{\tilde y}{-\xi + \pib {\trsp
          G}\sigma(\eta)}_d} %
    \mathcal{F}_\h\psi(\eta) \dd{\eta}\dd{\xi}\dd{\tilde{y}} \,.
  \end{equation}
  Using that
  \[
    \int e^{\frac{i}{\h} \pscal{\tilde y}{-\xi + \pib {\trsp
          G}\sigma(\eta)}_d} \dd {\tilde y} = \delta_{\{\xi = \pib
      {\trsp G} \sigma(\eta)\}}\,,
  \]
  we obtain
  \[
    U_\gamma U_G \psi (\tilde x) = \int e^{\frac{i}{\h}\pscal{\tilde X}{{\trsp G}
        \sigma(\eta)}} %
    e^{\frac{i}{\h}\pscal{\gamma}{\sigma(\pib {\trsp G}
        \sigma(\eta))}} %
    \mathcal{F}_\h\psi(\eta) \dd{\eta}
  \]
  If ${\trsp G} \sigma(\eta) \in \mathcal{H}_+$, we have
  $\sigma(\pib {\trsp G} \sigma(\eta)) = {\trsp G} \sigma(\eta)$, and
  the integrand becomes
  \[
    e^{\frac{i}{\h}\pscal{G(\tilde X + \gamma)}{\sigma(\eta)}}
    \mathcal{F}_\h\psi(\eta) \,,
  \]
  which is the formula involved in $U_{G\circ \tau_\gamma}\psi$. On
  the other hand, if ${\trsp G} \sigma(\eta) \in \mathcal{H}_-$, then
  $\sigma(\pib {\trsp G} \sigma(\eta)) = S_0({\trsp G} \sigma(\eta))$,
  where $S_0$ is the orthogonal symmetry with respect to
  $\mathcal{P}_0$. Hence
  \[
    \pscal{\gamma}{\sigma(\pib {\trsp G} \sigma(\eta))} =
    \pscal{S_0(\gamma)}{{\trsp G} \sigma(\eta)}\,,
  \]
  and therefore the integrand writes now
  \[
    e^{\frac{i}{\h}\pscal{G(\tilde X + S_0(\gamma))}{\sigma(\eta)}}
    \mathcal{F}_\h\psi(\eta) \,,
  \]
  which is the formula leading to $U_{G\circ \tau_{S(\gamma)}}\psi$.

\end{demo}
While the computation above is only formal, it gives rise to well
defined quantities under additional hypothesis, as discussed in
Section~\ref{sec:tilted-planes}: one may impose a compact support for
$\mathcal{F}_\h\psi$ inside $\{ \norm{\xi}^2 \leq 1\}$, or demand that
$\gamma$ and $G\tilde X$ belong to the right half-space
$\mathcal{H}_+$.

\subsection{Geometric study of $V_{\mathcal{G}}$}\label{sec:study-v_G}

Let $\mathcal{G}(\tilde X):= G\tilde X + \gamma$, with
$G\in\textup{GL}_{1+d}(\RM)$ and $\gamma\in\RM^{1+d}$.  We restrict
our study of $V_{\mathcal{G}}$~\eqref{equ:VG} to the region
$\norm{\xi}<1$, where the map $\sigma$ is smooth, and hence
$V_{\mathcal{G}}$ is a usual Fourier integral operator. The phase
function of $V_{\mathcal{G}}$ is
\begin{equation}
  \phi(\tilde x, \xi) = \pscal{\tilde{x}}{\pib({\trsp G}\sigma(\xi))}_d + \pscal{\gamma}{\sigma(\xi)},
  \label{equ:phase-VT}
\end{equation}
where, in view of the general discussion~\eqref{equ:oif},
$\tilde x \in \RM^d$, $\xi$ plays the role of $x$, and there is no
auxiliary variable. Hence it is automatically non-degenerate in the
sense of Hörmander, and $V_{\mathcal{G}}$ is a Fourier integral
operator associated with the canonical relation
$(\tilde x,\partial_{\tilde x}\phi,\, x, -\partial_x\phi)$. However,
in order to check whether it defines a (local) canonical
transformation, we should compute the mixed Hessian:
\begin{equation}
  \partial_{\tilde x}\partial_\xi\phi = \partial_{\xi}[\pib
  {\trsp G}\sigma(\xi)] = (\pib\circ {\trsp G}) \partial_{\xi}\sigma(\xi)\,.
  \label{equ:hessian}
\end{equation}
Notice that this Hessian is precisely the Jacobian matrix of the map
$g$ from Remark~\ref{rema:jacobian}. Thus, the Hessian is
non-degenerate if and only if $\trsp G\circ \sigma$ induces a local
diffeomorphism on the hyperplane $\mathcal{P}_0=\{0\}\times\RM^d$.
\begin{rema}
  Since $\tilde x$ and $\xi$ are multidimensional variables, it may be
  useful to explain some of the identifications that we use in this
  text. For any fixed $(\tilde x,\xi)$, we view
  $\partial_{\tilde x}\partial_\xi\phi$ as an endomorphism of $\RM^d$,
  with the usual implicit identifications. Precisely, if we view $\xi$
  as a covector (an element of the dual space $(\RM^d)^*$), then
  $\partial_{\tilde x}\partial_\xi\phi$ maps the tangent space
  $\Tg_x\RM^d\simeq \RM^d$ to $(\Tg_\xi(\RM^d)^*)^*\simeq \RM^d$. In
  other words, if
  $(u,v)\in \Tg_{(\tilde x,\xi)} (\Tg^*\RM^d) = \Tg_{\tilde x}\RM^d
  \times \Tg_\xi (\RM^d)^*$, we have
  \[
    \partial_{\tilde x}[\partial_\xi\phi (v)] (u) =
    \pscal{u}{(\pib\circ {\trsp G}) \dd{}_{\xi}\sigma\cdot v} =
    \partial_\xi [\partial_{\tilde x} \phi(u)](v)\,.
  \]
\end{rema}

With the following lemma, we first remark that the non-degeneracy is
easy to determine at $\xi=0$.
\begin{lemm}\label{lemm:orthogonal}
  In a neighbourhood of $\xi=0$, the Hessian~\eqref{equ:hessian} is
  non-degenerate if and only if the hyperplanes $P_0$ and $GP_0$ are
  not mutually perpendicular, \emph{i.e.}
  $\pscal{e_0}{G^{-1}e_0}\neq 0$.
\end{lemm}
\begin{demo}
  At $\xi=0$, the map $\sigma(\xi)$ is tangent to $(1,\xi)$, therefore
  $ \partial_{\xi}[\pib {\trsp G}\sigma(0)] = \pib{\trsp
    G}\pib$. Hence the Hessian is non-degenerate if and only if
  $\trsp G$ induces an isomorphism on the hyperplane
  $\mathcal{P}_0=\{0\}\times\RM^d$.  If $\pscal{e_0}{G^{-1}e_0}= 0$,
  the vector ${\trsp G}^{-1}e_0\in \mathcal{P}_0$ belongs to the
  kernel of $\pib\circ{\trsp G}$, which is hence not
  injective. Conversely, if there exists a non-zero
  $\Xi\in \mathcal{P}_0\cap\ker \pib\circ{\trsp G}$, then
  ${\trsp G}\Xi$ must be collinear to $e_0$, so $\Xi$ is collinear to
  ${\trsp G}^{-1}e_0$, which implies
  $\pscal{{\trsp G}^{-1}e_0}{e_0}=0$.
\end{demo}

In order to deal with the general position, we shall first give a
geometric argument, and then provide a precise algebraic computation
of the Hessian.

\begin{figure}[h]
  \centering \includegraphics[width=0.4\linewidth]{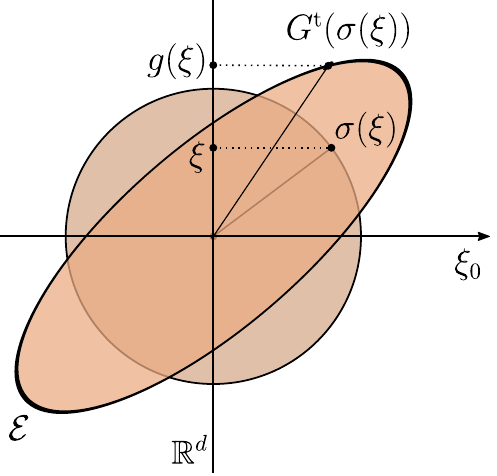}
  \caption{The map $g = \pib \circ {\trsp G} \circ \sigma$.}\label{fig:g}
\end{figure}

\begin{figure}[h]
  \centering \includegraphics[width=0.6\linewidth]{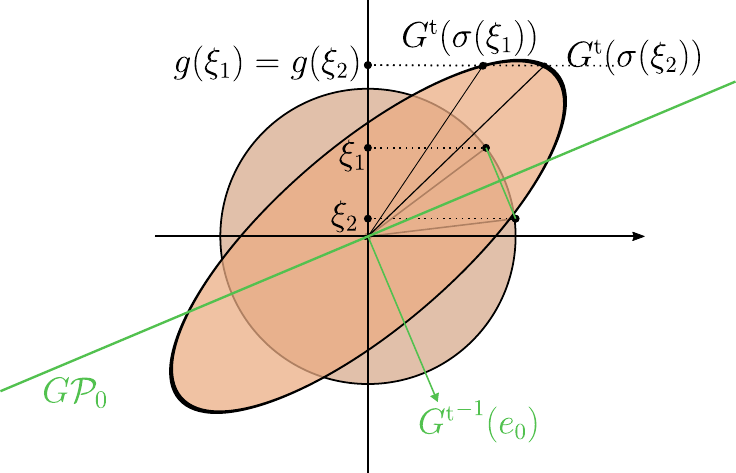}
  \caption{The map $g$ is not injective.}\label{fig:g2}
\end{figure}

Recall that
\[
  g(\xi) = \pib \circ {\trsp G} \circ \sigma (\xi).
\]
The map $\sigma$ is a diffeomorphism from the open unit ball
$\norm{\xi}<1$ to the right hemisphere in $\RM^{1+d}$. Applying
${\trsp G}$ on the vector $\sigma(\xi)$, we get a point on the Fourier
ellipsoid $\mathcal{E}={\trsp G}S^{d}$. The image $g(\xi)$ is the
projection of that point onto the hyperplane $\{\xi_0=0\}$, see
Figure~\ref{fig:g}.
This map is in general not injective:
\[
  \begin{aligned}
    g(\xi_1) = g(\xi_2) & \ssi \pib({\trsp G}\sigma(\xi_1) - {\trsp G}\sigma(\xi_2))  =0\\
                        & \ssi {\trsp G}\sigma(\xi_1) - {\trsp G}\sigma(\xi_2) = \lambda
                          e_0 \qquad \text{ for some } \lambda\in\RM\,.
  \end{aligned}
\]
Hence ${\trsp G}\sigma\xi_1$ and ${\trsp G}\sigma\xi_2$ are two points
on the ellipsoid that lie on a line parallel to $e_0$.  In general
position, a line intersects a quadric at 0 or 2 points; when we do
have two distinct points, the map $\pib_{\restr \mathcal{E}}$ is a
local diffeomorphism. At the critical points, where there is only one
intersection point (this is the ``apparent contour'' of the ellipsoid
in the direction $e_0$), $\pib_{\restr \mathcal{E}}$ cannot be a local
diffeomorphism. Since
$g = \pib_{\restr \mathcal{E}} \circ {\trsp G}\sigma$, its
differential will be degenerate exactly when
$\pib_{\restr \mathcal{E}}$ is not a local diffeomorphism.  Writing
\[
  g(\xi_1) = g(\xi_2) \ssi \sigma(\xi_1) - \sigma(\xi_2) = \lambda
  {\trsp G}^{-1} e_0 \qquad \text{ for some } \lambda\in\RM\,,
\]
we see that $\sigma(\xi_1)$ and $\sigma(\xi_2)$ are the intersection
points of the sphere $S^d$ with a line directed by
${\trsp G}^{-1} e_0$. (In other words, $\sigma(\xi_1)$ and
$\sigma(\xi_2)$ are images to each other under the orthogonal
reflection with respect to the hyperplane $G\mathcal{P}_0$). These two
points coalesce precisely when $\sigma(\xi_1)=\sigma(\xi_2)$ is
orthogonal to ${\trsp G}^{-1} e_0$. (See also Figure~\ref{fig:g2}.)
This proves the following.

\begin{prop}\label{prop:hessian}
  The Hessian~\eqref{equ:hessian} is degenerate (\emph{i.e.} non
  invertible) if and only if $\sigma(\xi)\in G \mathcal{P}_0$.

  Thus, microlocally near any $(x,\xi)\in T^*\RM^d$ such that
  $\norm{\xi}<1$, the operator $V_{\mathcal{G}}$ is a good Fourier
  integral operator with a non-degenerate phase function, and it is
  associated with a local symplectic diffeomorphism if and only if
  \begin{equation}
    \label{equ:non-degenerate}
    \sigma(\xi)\not\in G \mathcal{P}_0.
  \end{equation}
\end{prop}
\begin{rema}
  Physically, $\sigma(\xi)$ represents the direction of the light ray
  (see Section~\ref{sec:geometric-optics}), so the condition for a
  non-degenerate Hessian is that the light ray (in position space)
  must not be parallel to the rotated hyperplane $GP_0$; this is very
  natural: if this condition is not fulfilled, the impact of this
  light ray on the rotated hyperplane is not well defined.
\end{rema}

Let us now turn to a more computational argument. In order to make
Formula~\eqref{equ:hessian} more explicit, note that
$(\pib\circ {\trsp G})$ is the $d\times (1+d)$ matrix composed of the
last $d$ lines of ${\trsp G}$; we may write
\begin{equation}
  \pib\circ {\trsp G} =
  \begin{pmatrix}
    b & \trsp G_0
  \end{pmatrix}\label{equ:G0}
\end{equation}
where $b:=\pib\circ \trsp G (e_0) = g(0)\in\RM^{d}$, viewed as a
column vector, and $\trsp G_0$ is the lower-right $d\times d$ minor of
the matrix $\trsp G$. Let us denote
$f(\xi) := \sqrt{1-\snorm{\xi}^2}$.  The linear map
$\dd{}_{\xi}\sigma = (\dd f(\xi),\dd{\xi_1},\dots,\dd{\xi_d})$ is
represented by the $(1+ d)\times d$ matrix:
\[
  \dd{}_{\xi}\sigma =
  \begin{pmatrix}
    \frac{-\trsp \xi}{f(\xi)} \\
    \textup{I}_d
  \end{pmatrix}
\]
where $\textup{I}_d$ is the $d\times d$ identity matrix and
$\trsp \xi$ is the line vector $(\xi_1,\dots,\xi_d)$. Hence
\begin{equation}
  \dd g(\xi) =   (\pib\circ {\trsp G}) \dd{}_{\xi}\sigma =
  \frac{-b\cdot\trsp\xi}{f(\xi)} + \trsp G_0\,.
  \label{equ:hessian-formula}
\end{equation}
In particular, we recover the conclusion of
Lemma~\ref{lemm:orthogonal} that, if $\xi$ is small, the
non-degeneracy of the Hessian is implied by the invertibility of
$\trsp G_0$ (which is equivalent to the fact that the hyperplanes
$P_0$ and $GP_0$ are not orthogonal). In this case,
Formula~\eqref{equ:hessian-formula} gives
\begin{equation}
  \det [(\pib\circ {\trsp G}) \dd{}_{\xi}\sigma] = (\det {\trsp G_0})
  \left(1-\frac{\pscal{G_0^{-1}\xi}{b}}{f(\xi)}\right)
  \label{equ:hessian-determinant}
\end{equation}

\paragraph{Second proof of Proposition~\ref{prop:hessian}. }
Let $a:={\trsp G}^{-1}e_0$. Then $\sigma(\xi)\in G\mathcal{P}_0$ if
and only if $\pscal{\sigma(\xi)}a = 0$.

\paragraph{Case 1.} If $\trsp G_0$ is not invertible, then
$G\mathcal{P}_0$ is perpendicular to $\mathcal{P}_0$, \emph{i.e.}
$\pscal{e_0}a=0$, so $a\in \mathcal{P}_0$. Hence
$\trsp G_0 \pib a = \pib {\trsp G} a = \pib e_0 =0$, and on the other
hand $\pscal{\xi}{\pib a} = \pscal{\sigma(\xi)}a$. Therefore, if the
latter vanishes, $\pib a$ is a non-zero element of the kernel of the
Hessian~\eqref{equ:hessian-formula}, which is hence degenerate.

Conversely, if there is a non-zero $u\in\RM^d$ in the kernel of the
Hessian, \emph{i.e.}
\begin{equation}
  \pscal{\xi}u b = {\trsp G_0}u\,,
  \label{equ:kernel-hessian-orth}
\end{equation}
we can write this as
\[
  \pscal{\xi}u\pib {\trsp G}e_0 = \pib{\trsp G}\check u
\]
with $\check u := (0,u)\in \RM^{1+d}$. Therefore
$\pib{\trsp G}(\pscal{\xi}u e_0 - \check u)=0$, which means that
${\trsp G}(\pscal{\xi}u e_0 - \check u) = \lambda e_0$ for some
$\lambda\in\RM$, so
\[
  \pscal{\xi}u e_0 - \check u = \lambda {\trsp G}^{-1}e_0 = \lambda
  a\,.
\]
Taking the scalar product with $e_0$ we get
$\pscal{\xi}u = \lambda\pscal{a}{e_0} = 0$, which implies
by~\eqref{equ:kernel-hessian-orth} that ${\trsp G_0} u = 0$. But the
rank of $\trsp G_0$ must be $d-1$ (it is less than $d$ by assumption,
and it cannot be less than $d-1$ because $\pib\circ {\trsp G}$ must
have rank $d$). Hence, $u$ must be collinear with $\pib a$, an other
element of the kernel of ${\trsp G_0}$. Thus,
$\pscal{\sigma(\xi)}a = \pscal{\xi}{\pib a} = 0$.

\paragraph{Case 2.} We now assume that $\trsp G_0$ is invertible, so
$\pscal{a}{e_0}\neq 0$, and we may use
Formula~\eqref{equ:hessian-determinant}. Let us introduce the vector
$u:= {\trsp G_0}^{-1}b\in\RM^d$. The equality $b={\trsp G_0}u$ writes
\[
  \pib {\trsp G} e_0 = {\trsp G_0 u } = \pib {\trsp G} \check u\,,
\]
with $\check u := (0,u)\in\RM^{1+d}$. Therefore, there exists
$\lambda\in\RM$ such that ${\trsp G}(e_0 - \check u) = \lambda e_0$,
so $\check u = e_0 - \lambda {\trsp G}^{-1} e_0 = e_0 - \lambda a$. By
taking the scalar product with $e_0$ we get
\begin{equation}
  \label{equ:lambda}
  \lambda\pscal{a}{e_0} = 1\,,
\end{equation}
and by applying $\pib$ we get
\[
  u = - \lambda \pib a\,.
\]
Therefore,
\begin{gather}
  \pscal{\xi}u = -\lambda \pscal{\xi}{\pib a} = -\lambda
  \pscal{G^{-1}\check \xi}{e_0} = -\lambda \pscal{G^{-1}(\sigma(\xi) -
    f(\xi)e_0)}{e_0}\\
  = -\lambda \pscal{G^{-1}\sigma(\xi)}{e_0} +
  f(\xi)\lambda\pscal{e_0}a = -\lambda \pscal{G^{-1}\sigma(\xi)}{e_0}
  + f(\xi)\,.
\end{gather}
We see that the determinant~\eqref{equ:hessian-determinant} vanishes
if and only if $\pscal{\xi}u = f(\xi)$, \emph{i.e.} if and only if
$\pscal{G^{-1}\sigma(\xi)}{e_0} = 0$, which is the same as
$\pscal{\sigma(\xi)}a = 0$. \cqfd{}

Let us return to a more global viewpoint; the map
$\xi \mapsto {\trsp G} \sigma(\xi)$ sends the unit ball in $\RM^d$ to
a half ellipsoid $\mathcal{E}_r\subset \RM^{1+d}$. We can cut this
half ellipsoid into two pieces (separated by the apparent contour
$\mathcal{E}_0$ of the map $\pib$):
$\mathcal{E}_r = \mathcal{E}_- \sqcup \mathcal{E}_0 \sqcup
\mathcal{E}_+$, such that the map $\pib$ is injective on each
$\mathcal{E}_\pm$. Therefore, if we let
$D_j := \pib({\trsp G}^{-1}\mathcal{E}_j)$, then restricted to each of
the (semi-algebraic) open sets $D_-, D_+$, the map $g$ is now a smooth
diffeomorphism into its image $g(D_j) = \pib(\mathcal{E}_j)$. The
global lack of injectivity of $g$ on $B(0,1)$ is due to the fact that
the images $g(D_1)$ and $g(D_2)$ will in general overlap.

Equivalently, the hyperplane in $\RM^{1+d}$ normal to
$a={\trsp G}^{-1}e_0$ intersects the right hemisphere in $\RM^{1+d}$
into a $d-1$ dimensional hemisphere, whose projection $D_0$ onto
$\mathcal{P}_0$ separates the unit ball $\norm{\xi}<1$ into the two
sets $D_-,D_+$, that is
\begin{equation}
  B_{\mathcal{P}_0}(0,1) = D_- \sqcup D_0 \sqcup D_+\,,
  \label{equ:decomposition_D}
\end{equation}
with
\[
  D_\epsilon = \{\xi\in \RM^d; \norm{\xi}<1; \quad
  \pscal{\sigma(\xi)}a \in \epsilon(0,+\infty) \}\,, \quad
  \epsilon\in\{0,+,-\}\,.
\]
If $G=\textup{Id}$, then $D_- = \emptyset$.

\subsection{Study of $U_{\mathcal{G}}$ and canonical transformations}

When the condition of Proposition~\ref{prop:hessian} is fulfilled,
$V_{\mathcal{G}}$ is a microlocally invertible FIO associated with a
canonical transformation $\kappa_{V_{\mathcal{G}}} $ given implicitly
by
\begin{equation}
  \kappa_{V_{\mathcal{G}}} :   (\xi, -\partial_\xi\phi) \mapsto (\tilde x, \partial_{\tilde x}
  \phi).\label{equ:canonical-transform-V}
\end{equation}

Instead of the factorisation
$U_{\mathcal{G}} = V_{\mathcal{G}}\circ \mathcal{F}_\h$,
Formula~\eqref{equ:semiclassical-angular2} directly expresses the
operator $U_{\mathcal{G}}$ as a unique Fourier integral operator with
phase
\[
  \Phi(\tilde x, \xi, x) = \phi(\tilde x,\xi) - \pscal{x}\xi\,,
\]
where $\phi$ was defined in~\eqref{equ:phase-VT}. Physically speaking,
$U_{\mathcal{G}}$ is more relevant than $V_{\mathcal{G}}$, but of
course, it is computationally more involved, since the number of
integration variables has doubled with respect
to~\eqref{equ:semiclassical-angular1}. Let us compute the
corresponding canonical relation. In view of the general
theory~\eqref{equ:oif}, $\xi$ is the auxiliary variable, and we get
\[
  C_\Phi = \{(\tilde x, x, \xi); \, x = \partial_\xi\phi(\tilde x,
  \xi)\}
\]
which is always a smooth manifold of dimension $2d$. The canonical
relation $(\tilde x, \tilde \xi) = \kappa_{U_{\mathcal{G}}}(x,\xi)$ is
\[
  (\tilde x,\partial_{\tilde x}\Phi,\, x, -\partial_x\Phi)_{\restr
    C_\Phi} = (\tilde x, \partial_{\tilde x}\phi, \,
  \partial_\xi\phi(\tilde x,\xi), \xi) .
\]
As expected (in view of the fact that composition of FIOs is
associated with composition of canonical relations), this is the same
as the canonical transformation of~\eqref{equ:canonical-transform-V},
composed by the canonical transformation in the $(x,\xi)$ space
corresponding to the Fourier transform: $(x,\xi) \mapsto (\xi, -x)$:
\[
  (\tilde x, \tilde \xi) = \kappa_{U_{\mathcal{G}}}(x,\xi) =
  \kappa_{V_{\mathcal{G}}} (\xi,-x)\,.
\]

Despite the slightly more complicated phase function, it is
interesting to notice that the effective computation of the canonical
transformation $(\tilde x, \tilde \xi) = \kappa_\Phi(x,\xi)$
associated with $U_{\mathcal{G}}$ is actually easier than for
$V_{\mathcal{G}}$ alone. Indeed, first recall from~\eqref{equ:hessian}
that the map $g(\xi) = \pib({\trsp G}\sigma(\xi))$ has an invertible
differential, and that (from~\eqref{equ:phase-VT})
\[
  \phi(\tilde x, \xi) = \pscal{\tilde x}{g(\xi)}_d +
  \pscal{\gamma}{\sigma(\xi)}\,.
\]
Next, we see that $\tilde \xi = \partial_{\tilde x}\phi = g(\xi)$ does
not depend on $x$. This imposes the conjugate variable $\tilde x$ to
be obtained, up to the addition of a closed 1-form in $\tilde{\xi}$,
by the inverse adjoint of the differential of $g$:
\[
  \tilde x = {(\trsp {\dd{g}})}^{-1} x +
  [\text{closed}(\tilde{\xi})]\,.
\]
This is confirmed by the computation: let
$S(\xi) := \pscal{\gamma}{\sigma(\xi)}$; we have
$x=\partial_\xi \phi(\tilde x, \xi)$ and by~\eqref{equ:phase-VT}, for
all $v\in\RM^d$,
\[
  \partial_\xi \phi(\tilde x, \xi)\cdot v = \pscal{\tilde{x}}{\dd g\cdot v}_d
  + \dd{} S\cdot v \,,
\]
which gives $x = \trsp{(\dd{} g)} (\tilde x) + \dd S$. Locally, $g$
admits a unique inverse, which we denote by $g^{-1}$.  Let
$T = (g^{-1})^* S$, so that $\dd S= g^* \dd T$ and thus
\[
  \pscal{\dd S(\xi)}v = \pscal{\dd T(\tilde \xi)}{\dd{} g(\xi) \cdot
    v} = \pscal{\trsp{\dd{} g}(\xi) \dd T(\tilde \xi)}v\,.
\]
Hence we obtain the expected form
\begin{equation}
  \tilde x = {(\trsp{\dd g})}^{-1} x - {(\trsp{\dd g})}^{-1} \dd S =
  {(\trsp{\dd g})}^{-1} x - \dd T(g(\xi))\,.
  \label{equ:canonical-transform-U}
\end{equation}
This proves the formulas for the canonical transformations below.
\begin{theo}\label{theo:canonical} Let $g(\xi) = \pib({\trsp
    G}\sigma(\xi))$. The canonical relation of the Fourier integral
  operator $U_{G}$ (obtained when $\gamma=0$) in the region
  $\norm{\xi}<1$ is the set
  \[
    \{\left( x,\,g(\xi),\ {\trsp{\dd g}(\xi)}x,\,\xi\right), \quad
    x\in\RM^d, \xi\in \RM^d,\norm{\xi}<1 \}\,.
  \]
  Near any $\xi$ satisfying~\eqref{equ:non-degenerate}, this defines a
  canonical transformation $\kappa_{U_G}$ given by:
  \begin{equation}\label{equ:kappa_UG}
    \kappa_{U_G}(x,\xi) = \left({(\trsp{\dd g}(\xi))}^{-1} x,\,
      g(\xi)\right)\,.
  \end{equation}
  Let $S(\xi) = \pscal{\gamma}{\sigma(\xi)}$. The canonical
  transformation $\kappa_{U_\gamma}$ (obtained when $G=\textup{Id}$)
  is:
  \begin{align}
    \kappa_{U_\gamma}(x,\xi)
    & = (x - \dd S(\xi),\, \xi)\\
    & = \left( x + \frac{\gamma_0 \xi}{\sqrt{1-\norm{\xi}^2}} - \pib \gamma,\ \xi\right)\,,
      \label{equ:kappa-translation}
  \end{align}
  where $\gamma=(\gamma_0,\pib \gamma)\in \RM^{1+d}$.

  For a general affine transformation
  $\mathcal{G} = \tau_\gamma\circ G$, we
  have~\eqref{equ:canonical-transform-U}, which can be written:
  \begin{equation}
    \kappa_{U_{\mathcal{G}}} = \kappa_{U_G} \circ \kappa_{U_\gamma}\,.
    \label{equ:compose_kappa}
  \end{equation}
  Moreover, the map $\kappa_{U_\gamma}$ is a symplectomorphism from
  $\RM^d\times B(0,1)$ onto itself, while the map $\kappa_{U_G}$
  extends to two injective symplectomorphisms from $\RM^d\times D_\pm$
  (see~\eqref{equ:decomposition_D}), onto their respective images.
\end{theo}

The composition formula~\eqref{equ:compose_kappa} is in accordance
with Lemma~\ref{lemm:decomposition}.  In fact, another way to prove
Theorem~\ref{theo:canonical} is to use the
factorisation~\eqref{equ:matsushima-factorized}, which is valid only
in a microlocal sense:
\begin{equation}
  U_G = \mathcal{F}_\h^{-1}\circ J \circ (g^{-1})^* \circ
  \mathcal{F}_\h\label{equ:factorized-microlocal}
\end{equation}
which gives
\[
  U_{\mathcal{G}} = U_G \circ U_\gamma = \mathcal{F}_\h^{-1}\circ J
  \circ (g^{-1})^* \circ e^{\frac{i}{\h}S} \circ \mathcal{F}_\h\,;
\]
it then remains to compose the corresponding canonical
transformations, having in mind that multiplication by $J$ is a
pseudodifferential operator, so is associated with the identity
transformation.

\begin{figure}[h]
  \centering
  \includegraphics[width=0.9\linewidth]{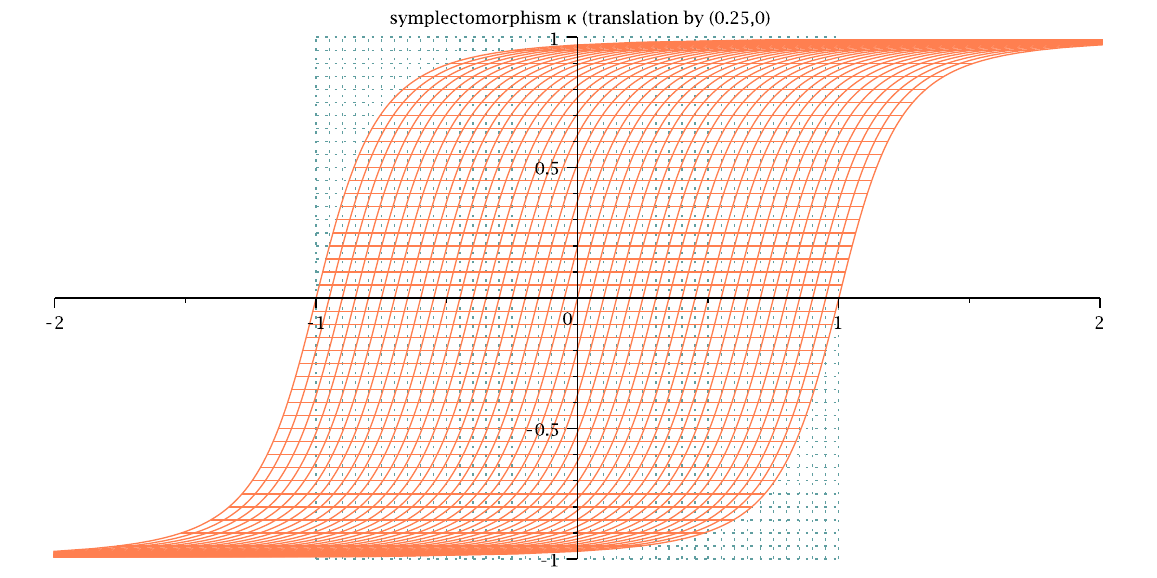}
  \caption{The symplectomorphism $\kappa_{U_\gamma}$, when
    $\gamma=(1/4,0)$. The dashed blue grid is the source domain
    $(-1,1)\times (-1,1)$ and the orange curves are its image under
    $\kappa_{U_\gamma}$. }\label{fig:translation_symplecto}
\end{figure}

\paragraph{The 1D case. ---}
The canonical transformations are quite explicit when $d=1$. For a
translation of vector $\gamma=(\gamma_0,\gamma_1)$,
formula~\eqref{equ:kappa-translation} becomes
\begin{equation}
  \label{equ:kappa-translation-1D}
  \kappa_{U_\gamma}(x,\xi) = \left(x+\frac{\gamma_0\xi}{\sqrt{1-\xi^2}} - \gamma_1,\ \xi\right), \qquad (x,\xi) \in \RM \times (-1,1)\,.
\end{equation}
See Figure~\ref{fig:translation_symplecto}.

\begin{figure}[h]
  \centering
  \includegraphics[width=0.5\linewidth]{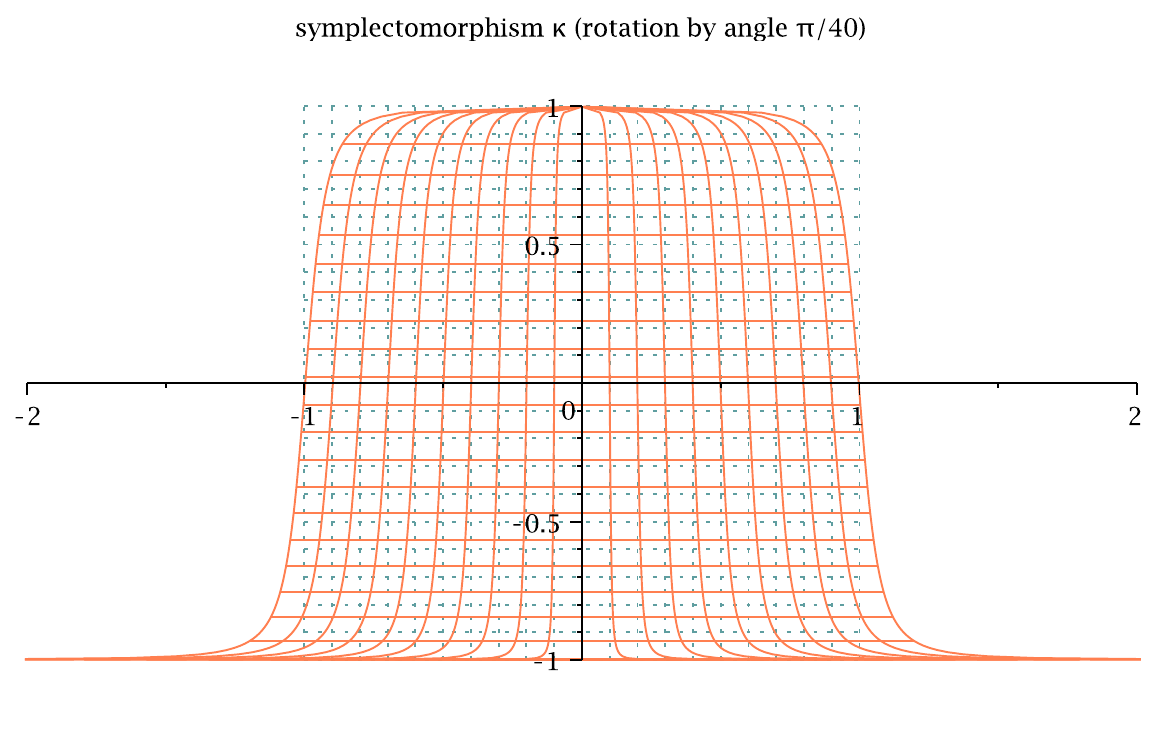}%
  \includegraphics[width=0.5\linewidth]{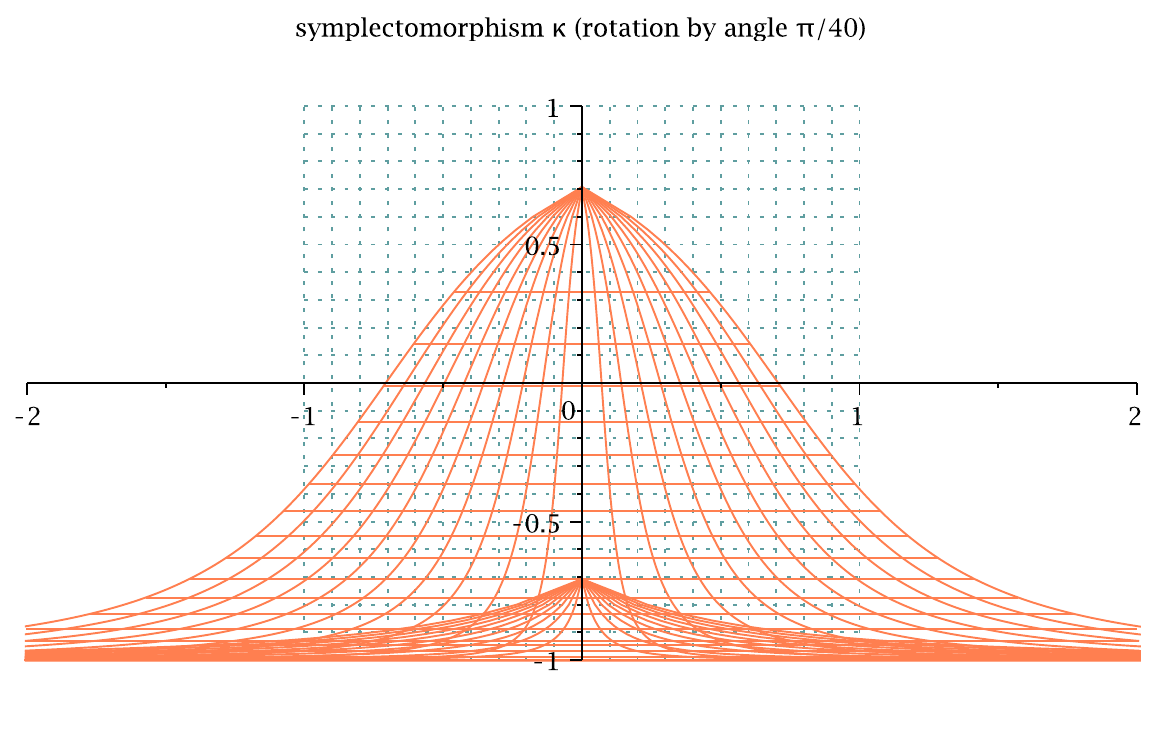}
  \caption{The symplectomorphism $\kappa_{U_G}$, when
    $G\in\textup{SO(2)}$ is the rotation of angle $\alpha=\pi/40$
    (left) and $\alpha=\pi/4$ (right). The dashed blue grid is the
    source domain $(-1,1)\times (-1,1)$ and the orange curves are its
    image under $\kappa_{U_G}$ (cropped to $(-2,2)$; the true image
    extends to $(-\infty,+\infty)\times{\{-1\}}$ when
    $\xi\to \xi_\alpha = -\cos(\alpha)$). On the right-hand side
    figure, we see the non-injectivity showing up at the bottom of the
    picture. }\label{fig:rotation_symplecto}
\end{figure}

For a rotation $G=
\begin{pmatrix}
  \cos\alpha & -\sin\alpha\\
  \sin\alpha & \cos\alpha
\end{pmatrix}
$ we have $g(\xi)= -\sqrt{1-\xi^2}\sin\alpha + \xi\cos\alpha$ and
hence~\eqref{equ:kappa_UG} becomes
\begin{equation}
  \label{equ:kappa_UG_1D}
  \kappa_{U_G}(x,\xi) = \left(\frac{x}{\frac{\xi\sin\alpha}{\sqrt{1-\xi^2}}+ \cos\alpha},\
    -\sqrt{1-\xi^2}\sin\alpha + \xi\cos\alpha\right)\,.
\end{equation}
See Figure~\ref{fig:rotation_symplecto}.

\begin{figure}[h!]
  \centering
  \includegraphics[width=0.5\textwidth]{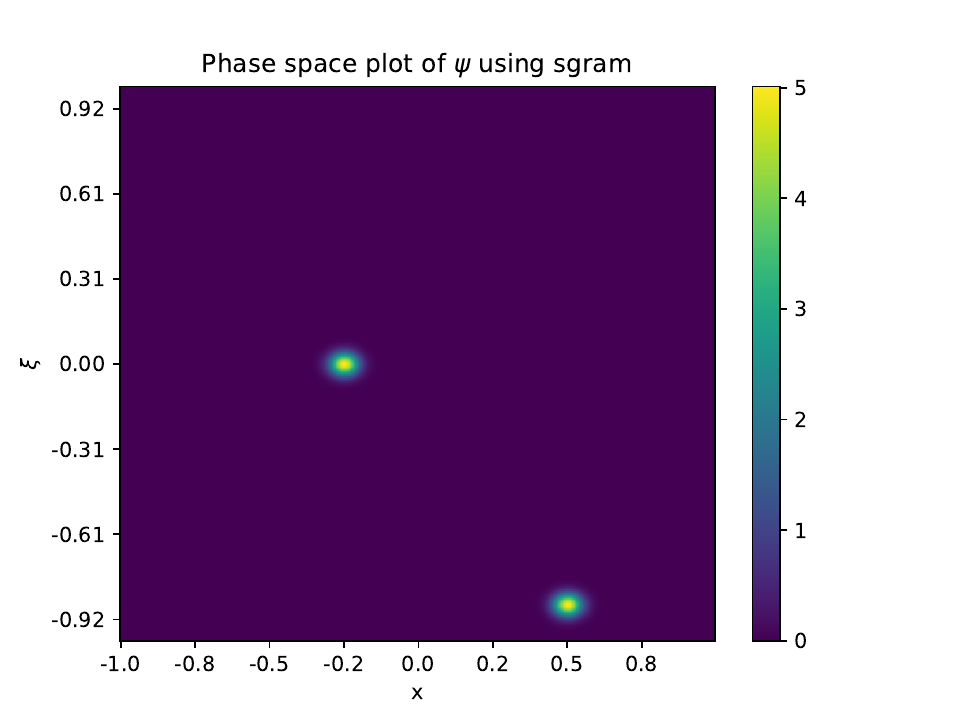}%
  \includegraphics[width=0.5\textwidth]{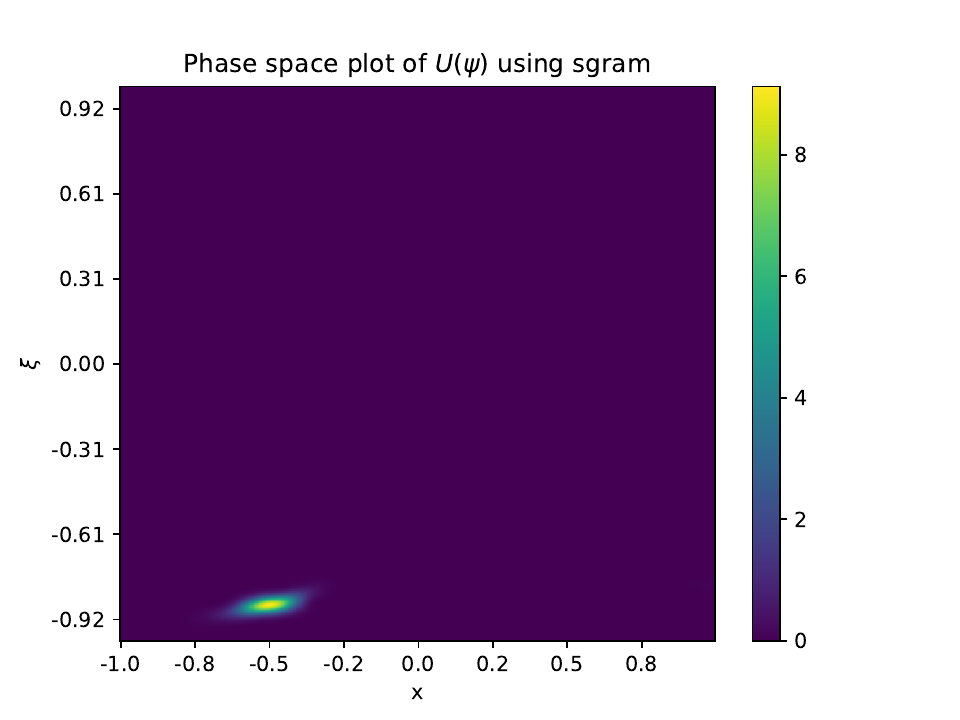}\\
  \includegraphics[width=0.5\textwidth]{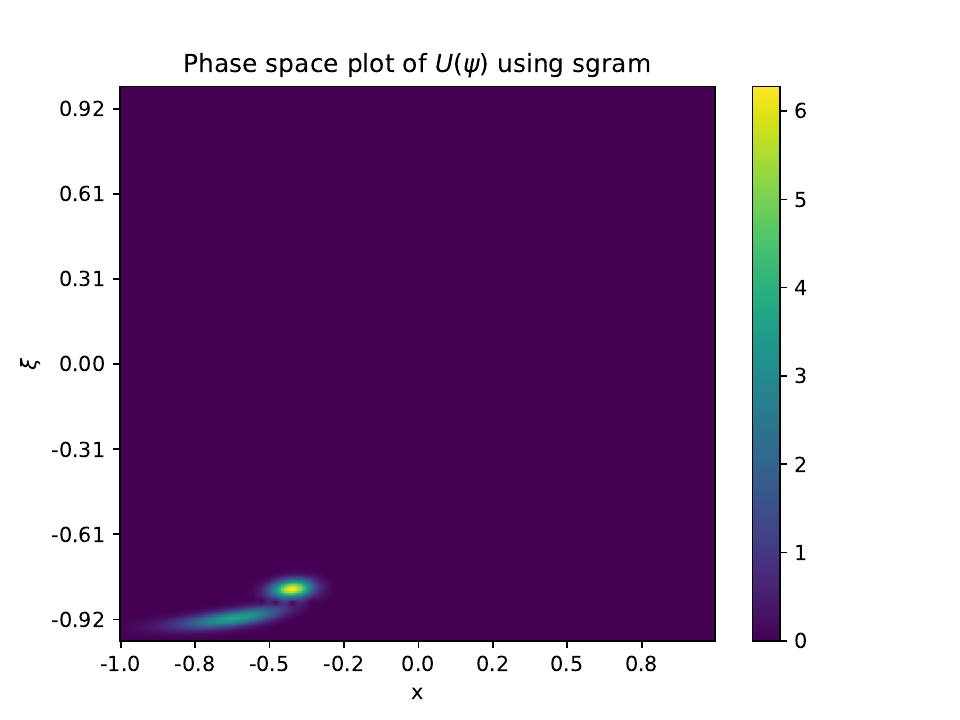}%
  \includegraphics[width=0.5\textwidth]{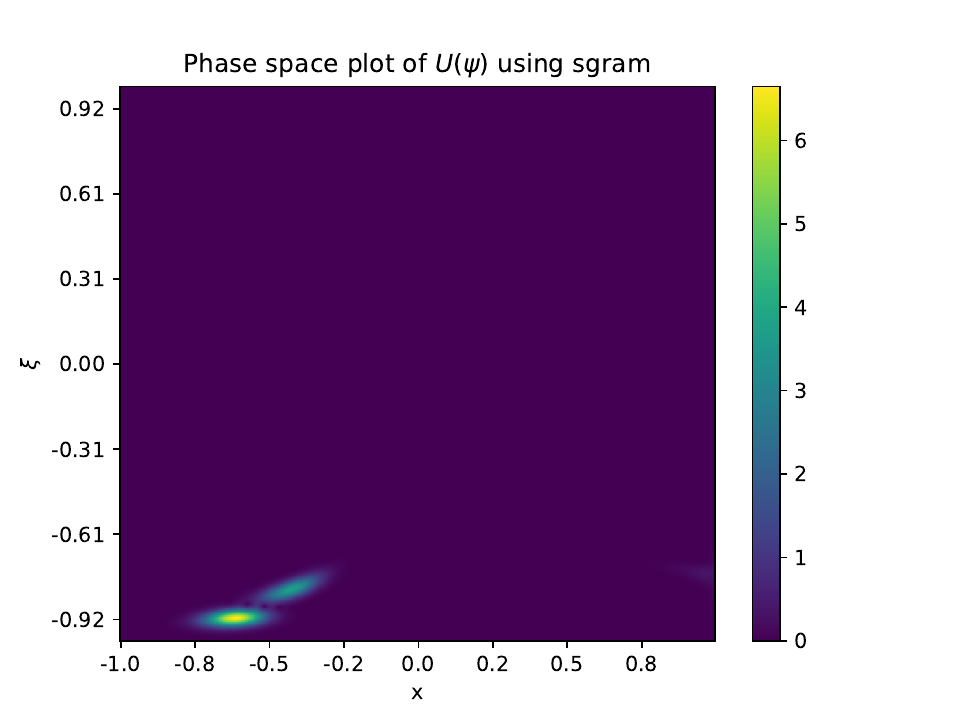}
  \captionsetup{singlelinecheck=off}
  \caption{The initial image (top left) is the phase space
    representation of a sum of two coherent states
    \[
      \psi(x) = e^{-\frac{(x-x_0)^2}{2\h}} e^{\frac{i}{\h}x\xi_0} + e^{-\frac{(x-x_1)^2}{2\h}} e^{\frac{i}{\h}x\xi_1}
    \]
    with $(x_0,\xi_0)=(-\frac{1}{4},0)$ and
    $(x_1,\xi_1)=(\frac{1}{2}, -\frac{\sqrt{3}}{2})$. It follows
    from~\eqref{equ:kappa_UG_1D} that they must superpose after a
    rotation $G$ of the plane $P_0$ by an angle
    $\alpha=\frac{\pi}{3}$. Hence, according to Egorov's theorem, the
    wave front set of $U_G\psi$ must contain only one point, which is
    confirmed by the numerical experiment (the top right image is a
    phase space representation of $U_G\psi$). The pictures below show
    the representations just before (bottom left, $\alpha=54^\circ$)
    and after (bottom right, $\alpha=66^\circ$) the ``collision''.  A
    movie showing the full rotation can be found
    online~\cite{san-nmovie}.}\label{fig:collision}
\end{figure}

Both formulas turn out to be even simpler if we switch to another set
of canonical variables
$(t,\theta)\in \RM\times(-\frac{\pi}{2},\frac{\pi}{2}) $ defined by
\[
  x =\frac{t}{\cos\theta}, \quad \xi = \sin\theta\,.
\]
Indeed, $g(\xi) = \sin (\theta-\alpha)$ and the formulas for the
corresponding canonical transformations
$\tilde\kappa(t,\theta) = (\tilde t, \tilde \theta)$ are
\[
  \tilde{\kappa}_{U_\gamma} (t,\theta) = (t + \gamma_0\sin\theta -
  \gamma_1\cos\theta,\ \theta)
\]
and, if $\theta-\alpha \in (-\frac{\pi}{2}, \frac{\pi}{2})$,
\[
  \tilde{\kappa}_{U_G}(t,\theta) =
  \begin{cases}
    (t,\ \theta-\alpha) & \text{ if } \theta-\alpha \in (-\frac{\pi}{2}, \frac{\pi}{2}) \mod 2\pi\\
    (-t,\ \pi+\alpha-\theta) &  \text{ if } \theta-\alpha\in (\frac{\pi}{2}, \frac{3\pi}{2}) \mod 2\pi\,.
  \end{cases}
\]
The non-degeneracy condition (Proposition~\ref{prop:hessian}) is here
simply $\theta-\alpha \neq \frac{\pi}{2} \mod \pi$, which means either
$\alpha=0$ modulo $\pi$, or $\xi\neq\xi_\alpha$ with
$\xi_\alpha:= -\varepsilon_\alpha\cos\alpha$ and
$\varepsilon_\alpha:=\textup{sign}(\sin\alpha)$. The injectivity
domains~\eqref{equ:decomposition_D} for $\xi$ are the intervals
$D_{-\varepsilon_\alpha}=(-1,\xi_\alpha)$,
$D_{\varepsilon_\alpha}=(\xi_\alpha, 1)$. For instance, when
$\alpha=\frac{\pi}{3}$, the points $(-\frac{1}{4}, 0)$ and
$(x_2=1/2, \xi_2=-\sqrt{3}/2)$ have the same image under
$\kappa_{U_G}$, see Figure~\ref{fig:collision}.

\subsection{Geometric optics}\label{sec:geometric-optics}

Based on the canonical transformations of
Theorem~\ref{theo:canonical}, which we obtained directly from the
solution to the Helmholtz equation, we explain in this section how to
recover the expected laws of geometric optics.

\begin{lemm}\label{lemm:classical}
  Let $(x_B,\xi_B) = \kappa_{U_{\mathcal{G}}}(x_A,\xi_A)$. We consider
  the usual inclusion ${\trsp \pib} : \RM^{d} \to \RM^{1+d}$ defined
  by ${\trsp \pib}(x) = (0,x)$. Let $A = {\trsp \pib} x_A$ and
  $B = \mathcal{G}{\trsp \pib} x_B$. Then, $B-A$ is collinear to
  $\sigma(\xi_A)$.
\end{lemm}
\begin{demo}
  Because of Theorem~\ref{theo:canonical}, we may treat the cases
  $\mathcal{G}=G$ and $\mathcal{G}=\tau_\gamma$ separately.

  \paragraph{Case $\mathcal{G}=G$.} In view of the notation
  from~\eqref{equ:G0}, this gives
  \[
    B = G{\trsp \pib} x_B =
    \begin{pmatrix} {\trsp b}\\G_0
    \end{pmatrix} x_B = \pscal{b}{x_B}e_0 + {\trsp \pib} G_0 x_B\,.
  \]
  From~\eqref{equ:kappa_UG}, $x_A = (\trsp{\dd g(\xi_A)}) x_B$, and
  using~\eqref{equ:hessian-formula}, we have
  \begin{align}
    x_A & = -\frac{\pscal{b}{x_B}}{f(\xi_A)}\xi_A + G_0 x_B\\
    \text{hence } A
        & = -\pscal{b}{x_B}\left(\frac{ {\trsp \pib}\xi_A + f(\xi_A)e_0}{f(\xi_A)}\right) + G{\trsp \pib} x_B \\
        & = -\frac{-\pscal{b}{x_B}}{f(\xi_A)}\sigma(\xi_A) + B
  \end{align}
  which proves the lemma for $G$.

  \paragraph{Case $\mathcal{G}=\tau_\gamma$.} We have
  $A={\trsp\pib x_A}$ and $B={\trsp \pib}x_B + \gamma$. From
  Theorem~\ref{theo:canonical}, $x_B = x_A - \dd S(\xi_A)$, so
  $B-A = \gamma - {\trsp\pib \dd S(\xi_A)}$. Consider the map
  $\xi\mapsto S(\xi) - \pscal{\pib \gamma}{\xi} =
  \pscal{\gamma}{\sigma(\xi)-{\trsp \pib}\xi} =
  f(\xi)\pscal{\gamma}{e_0}$. It differential
  $\dd S(\xi) - \pib \gamma$ is hence equal to
  $-\frac{\pscal{\gamma}{e_0}}{f(\xi)}\xi$. Hence
  \begin{align*}
    A - B =  {\trsp\pib}\dd S(\xi_A) - \gamma
    &= {\trsp\pib}(\dd S(\xi_A) - \pib \gamma) - \pscal{\gamma}{e_0}e_0\\
    &= -\frac{\pscal{\gamma}{e_0}}{f(\xi_A)}({\trsp \pib}\xi_A + f(\xi_A)e_0) \\
    &=  -\frac{\pscal{\gamma}{e_0}}{f(\xi_A)}\sigma (\xi_A)\,,
  \end{align*}
  which finishes the proof.
\end{demo}
Thanks to this lemma, the map $\kappa_{U_{\mathcal{G}}}$ can be
described as follows. To any point $(x_A,\xi_A)\in \RM^{d + d}$, we
associate the ``light ray'' in $\RM^{1+d}$ emanating from the point
$A={\trsp\pib} x_A = (0,x_A)\in P_0$ with direction
$\sigma(\xi_A)$. Under the condition of
Proposition~\ref{prop:hessian}, this line intersects the hyperplane
$\mathcal{G} P_0$ in a unique point $B$.  Let Point $B$ be
parameterised by $x_B\in \RM^d$ via
$B = \mathcal{G}{\trsp\pib} x_B = \mathcal{G}((0,x_B))$. It follows
from Lemma~\ref{lemm:classical} that the map $x_A\mapsto x_B$ is
exactly the $x$-component of the canonical transformation of
Theorem~\ref{theo:canonical}.

It remains to interpret the transformation of the $\xi$ component,
namely $\xi_B = g(\xi_A)$. For this we simply consider a plane wave on
$\RM^{1+d}$ directed by $\sigma(\xi_A)$:
\[
  \psi(X)=e^{\frac{i}{\h}\pscal{X}{\sigma(\xi_A)}}.
\]
It is a solution to the Helmholtz/Laplace/Schrödinger
equation~\eqref{equ:schrodinger}. Its restriction to $P_0$ is
$\psi_0 (x)= {(\trsp\pib)}^*\psi (x) =
e^{\frac{i}{\h}\pscal{x}{\xi_A}}$, for any $x\in\RM^d$: it is a plane
wave directed by $\xi_A$ (that is, with wave front set
$\textup{WF}_\h(\psi_0) = \RM^d \times \{ \xi_A\}$). We now consider
the transformed hologram
$\psi_1 = {(\trsp\pib)}^* (\mathcal{G}^*\psi) = (\mathcal{G} \circ
{\trsp\pib})^*\psi$. First of all, since $\kappa_{U_\gamma}$ preserves
the $\xi$ component, we may assume that
$\mathcal{G}=G\in \textup{GL}_{1+d}(\RM)$. We then compute easily
$\psi_1(x) = e^{\frac{i}{\h}\pscal{x}{\pib {\trsp G} \sigma(\xi_A)}} =
e^{\frac{i}{\h}\pscal{x}{g(\xi_A)}} $, so $\psi_1$ is again plane wave
directed by $\xi_B = g(\xi_A)$, as expected. Note also that, if
${\trsp G}\sigma(\xi_A)$ belongs to the right half-space, then this
relation is equivalent to $\sigma(\xi_B) = {\trsp G}\sigma(\xi_A)$,
which expresses the new direction of the light ray, viewed in the
transformed hologram, in terms of its direction in the original
hologram.

\begin{rema}
  By construction, the canonical transformations of
  Theorem~\ref{theo:canonical} are symplectic with respect to the
  usual (canonical) symplectic form $\dd\xi \wedge \dd x$ of
  $T^*P_0$. One may wonder whether this symplectic form is natural,
  given the fact that the original phase space of our problem is
  rather $T^*\RM^{1+d}$, and $P_0$ is just an embedded hyperplane. A
  way to see this is to consider the restriction operator
  ${(\trsp \pib)}^*: \psi \mapsto \psi_{\restr P_0} = x \mapsto
  \psi({\trsp \pib}(x))$, which we implicitly use to define
  $U_{\mathcal{G}}$. It can be written as the Fourier integral
  operator
  \[ {(\trsp \pib)}^* \psi (x)= \frac{1}{(2\pi\h)^{1+d}} \int
    e^{\frac{i}{h}(\pscal{\trsp
        \pib(x)}{\theta}-\pscal{X}{\theta})}\psi(X)\dd{X}\dd{\theta}\,.
  \]
  (The difference with~\eqref{equ:oif} is that we now allow $X$ and
  $x$ (which play the roles of $x$ and $\tilde x$ there, respectively)
  to have different dimensions, but the theory remains valid.) The
  phase function is
  $\phi(x, X, \theta) = \pscal{\trsp
    \pib(x)}{\theta}-\pscal{X}{\theta}$, the manifold $C_\phi$ is
  given by $X={\trsp \pib}(x)$, and hence the Lagrangian
  $\Lambda_\phi$ is
  \[
    \Lambda_\phi = \{(x, \dd{\pib}\cdot \theta, {\trsp \pib}(x),
    \theta), \quad x\in \RM^{d}, \theta \in (\RM^{1+d})^* \}\,.
  \]
  If we write $\theta=(\theta_0, \xi)$ with $\xi\in (\RM^{d})^*$ (so
  that $\dd{\pib}\cdot \theta = \xi$), we see that this canonical
  relation sends the restriction of the canonical symplectic form
  $(\dd \theta \wedge \dd X)_{\restr P_0\times (\RM^{1+d})^*}$ to
  $\dd{\xi} \wedge \dd{x}$, which shows that the latter is the correct
  symplectic form to consider on $T^*P_0$.

  Thus, the symplectic form $\dd\xi \wedge \dd x$ is dictated by the
  choice of the Angular Spectrum formula. Its simplicity makes it an
  appealing choice for microlocal formulas; however, as mentioned
  before, it raises difficulties due to the non-injectivity of the
  canonical transformation. It could be interesting to develop a
  microlocal analysis of holograms directly on the co-sphere, as
  developed in a different context by~\cite{cheverry-ibrahim25}.

\end{rema}

\section{Precised Egorov Theorem}\label{sec:prec-egor-theor}

\begin{figure}[h!]
  \centering
  \includegraphics[height=5cm]{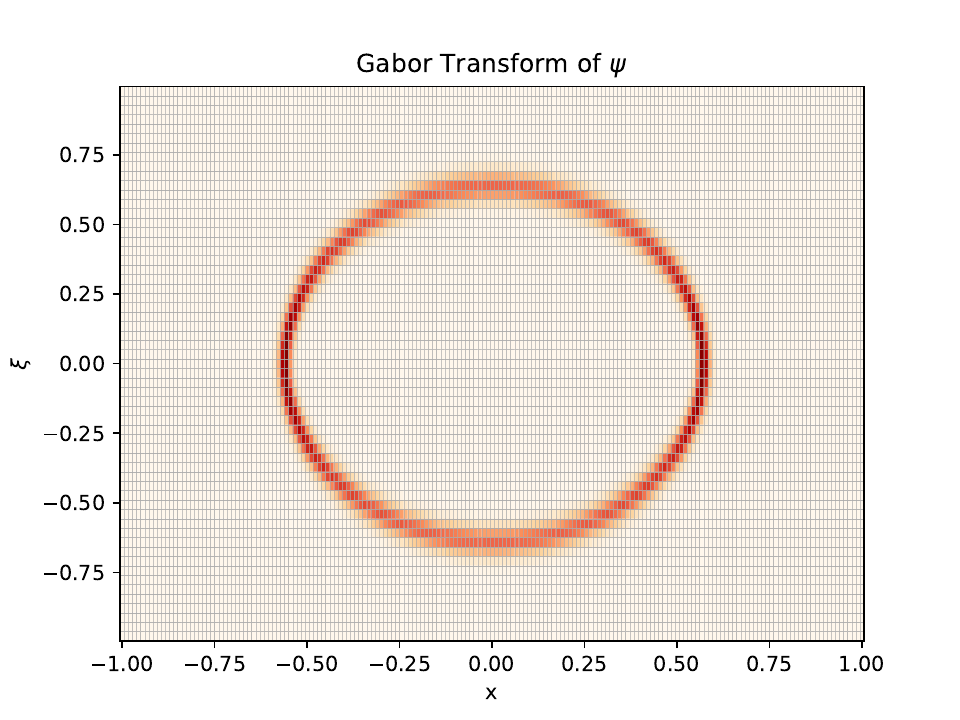}\\
  \includegraphics[height=5cm]{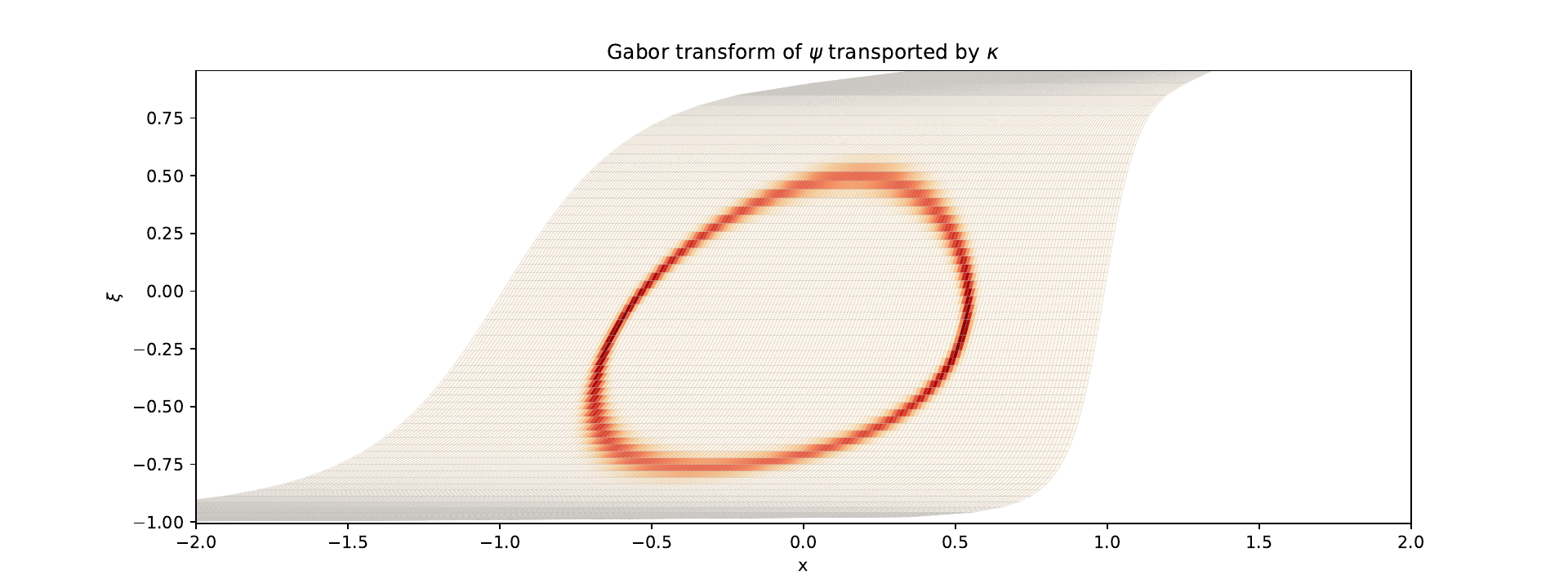}\\
  \includegraphics[height=5cm]{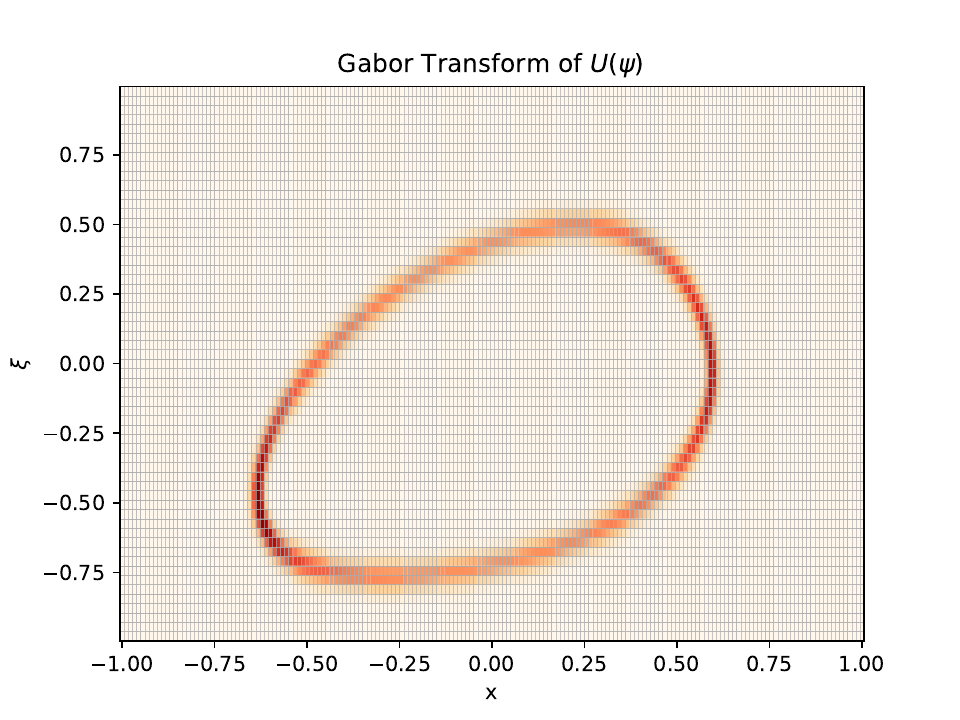}

  \caption{Here $\psi$ is a Hermite function (same as
    Figure~\ref{fig:hermite}). The first figure shows its
    semiclassical wavefront set, using a semiclassical Gabor
    transform. The second figure is obtained simply by applying the
    canonical transformations of Theorem~\ref{theo:canonical} to
    Figure 1, with $\mathcal{G}X= GX + \gamma$, where $G$ is a
    rotation of angle 10° and $\gamma=(\frac{3}{10},0)$. It should be
    compared to the last figure, which is the semiclassical wavefront
    set of the propagated signal $U_{\mathcal{G}}(\psi)$: one can see
    that the curves coincide quite perfectly, which is predicted by
    Egorov's theorem.}\label{fig:hermite-egorov}
\end{figure}

The canonical transformation of Theorem~\ref{theo:canonical} gives a
first approximation of the propagator $U_{\mathcal{G}}$ in the sense
of geometric optics: if a signal $\psi$ is localised in phase space in
some region $\Omega$ (its \emph{wave front set} is contained in
$\Omega$), then $U_{\mathcal{G}}\psi$ is localised in
$\kappa(\Omega)$, where $\kappa$ is the canonical transformation
associated with $U_{\mathcal{G}}$, see
Figure~\ref{fig:hermite-egorov}. However, the notion of wave front set
is not very precise: its hides the information about how $\psi$
concentrates on the classical rays, and about the phase of $\psi$.

In order to obtain precise information about the propagated signal
$U_{\mathcal{G}}\psi$, we need to introduce the notion of
\emph{quantum observables} (in signal processing, they would be
time-frequency filters --- although here the phase space is rather
``position-direction''). They are selfadjoint operators $P$, and the
``observation'' of a normalised state (or signal) $\Psi$ by $P$ is by
definition the scalar product $\pscal{P \Psi}{\Psi}$. For instance,
the operator $P$ can be the position operator $P=x_j$, the momentum
operator $P=\frac{\h}{i}\partial_{x_j}$, or any combination, that is,
a pseudodifferential operator with symbol $p(x,\xi)$. If we wish to
observe the propagated signal $U_{\mathcal{G}}\psi$, we are led to
compute
\[
  \pscal{P U_{\mathcal{G}}\psi}{U_{\mathcal{G}}\psi} =
  \pscal{U_{\mathcal{G}}^* P U_{\mathcal{G}}\psi}{\psi}\,.
\]
Thus, we see that this amounts to the observation of the
\emph{initial} state $\psi$ with the new operator
$U_{\mathcal{G}}^* P U_{\mathcal{G}}$. The main tool to compute this
new operator is the Egorov Theorem, whose general statement is as
follows. Let $P$ be a pseudodifferential operator with principal
symbol $p(x,\xi)$. Then, when $U_{\mathcal{G}}$ is microlocally
invertible,
\[
  U_{\mathcal{G}}^{-1} P U_{\mathcal{G}} = R
\]
where $R$ is a pseudodifferential operator with principal symbol
$r = p \circ \kappa_{U_{\mathcal{G}}}$. If we fix a quantization
scheme, for instance Weyl quantization, which associates to a symbol
$p(x,\xi)$ a pseudodifferential operator $P=\Op_\h^w(p)$, we obtain

\[
  U_{\mathcal{G}}^{-1} \Op_\h^w(p) U_{\mathcal{G}} =
  \Op_\h^w(p\circ\kappa_{U_{\mathcal{G}}}) + \O(\h)
\]
Suppose first that $U_{\mathcal{G}}$ is unitary:
$U_{\mathcal{G}}^{-1} = U_{\mathcal{G}}^*$. Then, in order to observe
the propagated signal $U_{\mathcal{G}}\psi$ with $ \Op_\h^w(p)$,
Egorov's theorem tells us that it is enough to observe the initial
signal $\psi$ with the operator
$\Op_\h^w(p\circ\kappa_{U_{\mathcal{G}}})$, up to small errors of
order $\O(\h)$. This is appealing for applications, since we can avoid
computing the exact propagation $U_{\mathcal{G}}$, and content
ourselves with the classical canonical transformation
$\kappa_{U_{\mathcal{G}}}$.

However, if we really want to observe effects that go beyond geometric
optics, we should understand the $\O(\h)$ remainder. This term is
notoriously more difficult to compute, as it depends on the precise
(non-geometric) formula for the quantum propagator. In this work, we
obtain an explicit formula for this remainder, given in
Theorem~\ref{theo:egorov} which, to the best of our knowledge, is a
new result.

Finally, in our case, $U_{\mathcal{G}}$ is not unitary, and the
formula for $U_{\mathcal{G}}^{-1} P U_{\mathcal{G}}$ does not apply to
$U_{\mathcal{G}}^* P U_{\mathcal{G}}$. In order to obtain the latter,
we need to compute the defect of unitarity, which we perform in
Section~\ref{sec:lack-unitarity}. As a result, this proves that the
Egorov theorem for $U_{\mathcal{G}}^* P U_{\mathcal{G}}$ holds with
accuracy $\O(\h^2)$ (Theorem~\ref{theo:egorov-adjoint}).

In the case of the Schrodinger equation (which can be seen as an
approximation of Helmholtz' equation in the paraxial
regime~\cite{doumic-2003}), related Egorov-type theorems have been
obtained, see for instance~\cite{gaim-2014} and references therein. In
these works, the fact that the propagator is really the exponential of
a (pseudo)differential simplifies the computation of higher order
terms. Unfortunately, we can use these techniques for our Helmholtz
propagator.

\begin{rema}
  In the case of (microlocally) unitary Fourier integral operators $U$
  of the form~\eqref{equ:oif}, it was proved
  in~\cite{hitrik-sjostrand-I} that Egorov's theorem holds with
  $\O(\h^2)$ remainder when the phase of $a_0$ is constant. It could
  be interesting to try to use their result to give a different proof
  of Theorem~\ref{theo:egorov-adjoint}.
\end{rema}

\subsection{Products of $V_G$ with pseudodifferential operators}

Phase space filters are essential for holographic studies because they
allow (smooth) truncation simultaneously in position and
direction. For instance, the eye is naturally such a phase space
filter, for not only has it a specific position is space, but also it
selects a narrow beams of light rays which are directed towards it. In
our phase space analysis, filter are conveniently represented by
semiclassical pseudodifferential operators. Our goal in this section
is to study the two cases where the filter is applied before or after
the propagation $U_{\mathcal{G}}$. Since the action of the Fourier
transform is easy to deal with, and in view of the decomposition given
by Lemma~\ref{lemm:decomposition}, it is enough to consider products
of $V_G$, $G\in \textup{GL}_{1+d}(\RM)$, with pseudodifferential
operators, where $V_G$ is the integral operator defined
in~\eqref{equ:VG}.

Let $P,Q$ be semiclassical pseudodifferential operators. We consider
$PV_G$ and $V_G Q$. In terms of integral (\emph{i.e.} Schwartz)
kernels, we have
\[
  K_{V_G}(y, \eta) = \frac{ e^{\frac{i}{\h}\phi(y, \eta)}}{(2\pi\h)^d}
\]
\[
  K_P(x,y) = \int_{\RM^d}e^{\frac{i}{\h}\pscal{x-y}{\tilde\xi}}
  p(\tfrac{x+y}{2}, \tilde \xi) \dd{\tilde{\xi}}
\]
\[
  K_Q(\eta,\xi) = \int_{\RM^d}e^{\frac{i}{\h}\pscal{\eta-\xi}{\tilde x}}
  q(\tfrac{\eta+\xi}{2}, \tilde x)\dd{\tilde x}
\]
Hence
\[
  K_{PV_G}(x,\eta) = \frac{1}{(2\pi\h)^d}\intint
  e^{\frac{i}{\h}\pscal{x-y}{\tilde \xi}} p(\tfrac{x+y}{2}, \tilde \xi)
  e^{\frac{i}{\h}\phi(y, \eta)}\dd{\tilde \xi} \dd y
\]
and
\[
  K_{V_G Q}(y, \xi) = \frac{1}{(2\pi\h)^d}\intint
  e^{\frac{i}{\h}\pscal{\eta-\xi}{\tilde x}} q(\tfrac{\eta+\xi}{2},
  \tilde x) e^{\frac{i}{\h}\phi(y, \eta)} \dd{\eta}\dd{\tilde x} \,.
\]
These are oscillatory integrals with respective phases
\[
  \Phi^{PV_G}_{(x,\eta)}(\tilde \xi, y) = \pscal{x-y}{\tilde \xi} +
  \phi(y, \eta) = \pscal{x-y}{\tilde \xi} + \pscal{y}{\pib({\trsp
      G}\sigma(\eta))},
\]
and
\[
  \Phi^{V_G Q}_{(y,\xi)}(\eta, \tilde x) = \pscal{\eta-\xi}{\tilde x} +
  \phi(y, \eta) = \pscal{\eta-\xi}{\tilde x} + \pscal{y}{\pib({\trsp
      G}\sigma(\eta))} \,.
\]

The first one, $\Phi^{PV_G}_{(x,\eta)}(\tilde \xi, y)$, is associated
with the Lagrangian manifold (its wavefront) $\Lambda_{PV_G}$ given by
\[
  \partial_{\tilde \xi}\Phi^{PV_G}_{(x,\eta)} = 0; \quad
  \partial_{y}\Phi^{PV_G}_{(x,\eta)} = 0\,.
\]
This gives $x=y$ and
$\tilde\xi = \partial_y\phi(y,\eta) = \pib({\trsp G}\sigma(\eta)) =
g(\eta)$, and the value of the phase on $\Lambda_{PV_G}$ is simply
$\phi(x,\eta)$.  Notice that $\Phi^{PV_G}_{(x,\eta)}(\tilde \xi, y)$
is polynomial of degree 2 in its variables $(\tilde\xi,y)$. Hence the
stationary phase formula is ``explicit'' and yields (see for
instance~\cite[Theorem 15.5.1]{guillemin-sternberg-semiclassical}
or~\cite[Theorem 3.13]{zworski-book-12})
\[
  K_{PV_G}(x,\eta) \sim e^{\frac{i}{\h}\phi(x,\eta)}
  \frac{e^{i\frac{\pi}{4}\textup{sgn}(\mathcal{Q}(x,\eta))}}{\abs{\det
      \mathcal{Q}(x,\eta)}^{1/2}} \sum_{k\geq 0}\frac{\h^k}{k!}
  \left(\frac{\pscal{\mathcal{Q}(x,\eta)^{-1}D}{D}}{2i}\right)^k
  p(\tfrac{x+y}{2},\tilde \xi)
\]
taken at $y=x$, $\tilde\xi = \pib({\trsp G}\sigma(\eta))$. We have
denoted the quadratic form
$\mathcal{Q}(x,\eta) := (\Phi^{PV_G}_{(x,\eta)})''(0)$ and
$D=\frac{1}{i}(\partial_{\tilde \xi},\partial_{y})$.

Since the quadratic part of $\Phi^{PV_G}_{(x,\eta)}(\tilde \xi, y)$ is
simply $- \pscal{y}{\tilde \xi}$, its eigenvalues are
$(1,\dots,1,-1,\dots,-1)$ and hence the signature vanishes and the
determinant is 1; moreover $\mathcal{Q}^{-1} = \mathcal{Q}$, so
\[
  \pscal{\mathcal{Q}(x,\eta)^{-1}D}{D} =
  2\pscal{\partial_{\tilde\xi}}{\partial_y}
\]
and
\begin{align}
  K_{PV_G}(x,\eta)
  & \sim e^{\frac{i}{\h}\pscal{x}{\tilde\xi}}
    \sum_{k\geq 0}\frac{(-i\h)^k}{k!}
    \pscal{\partial_{\tilde \xi}}{\partial_y}^k
    p(\tfrac{x+y}{2},\tilde \xi)_{y=x} \nonumber\\
  & \sim e^{\frac{i}{\h}\pscal{x}{\tilde\xi}} \exp\left(\frac{\h}{2i}\pscal{\partial_{\tilde \xi}}{\partial_x}\right) p(x,\tilde \xi) \nonumber\\
  & =   e^{\frac{i}{\h}\pscal{x}{\tilde\xi}}
    \left(p(x,\tilde\xi) -
    \frac{i\h}{2}\pscal{\partial_{\tilde \xi}}{\partial_x}p(x,\tilde \xi) + \O(\h^2)\right)
\end{align}
Of course, this works equally well when $p=p_\h$ admits an asymptotic
expansion of the form $p_\h = p_0 + \h p_1 + \O(\h^2)$. We obtain
\begin{prop}\label{prop:KPV}
  Let $P=\Op_\h^w(p_\h)$, with a symbol $p_h$ of the form
  $p_\h = p_0 + \h p_1 + \O(\h^2)$. We have, in the $\Cinf$ topology,
  \begin{equation}
    \label{equ:KPVT}
    K_{PV_G}(x,\eta) =  e^{\frac{i}{\h}\pscal{x}{\tilde\xi}}
    \left(p_0(x,\tilde\xi) + \h p_1(x,\tilde\xi) -
      \frac{i\h}{2}\pscal{\partial_{\tilde \xi}}{\partial_x}p_0(x,\tilde \xi) + \O(\h^2)\right)\,,
  \end{equation}
  where $\tilde{\xi} = g(\eta)$.
\end{prop}

The second composition, $V_G Q$, is more complicated because the phase
is not quadratic anymore. The Lagrangian manifold $\Lambda_{V_G Q}$ is
given by
\[
  \partial_{\tilde x} \Phi^{V_G Q}_{(y,\xi)}(\eta, \tilde x) = 0 \quad
  \partial_{\eta} \Phi^{V_G Q}_{(y,\xi)}(\eta, \tilde x) =0\,,
\]
which gives $\eta=\xi$ and
$\tilde x = -\partial_\eta\phi(y,\eta) = - {\trsp {\dd g}}(\eta)\cdot
y$.  We will now Taylor expand the phase $\phi(y, \eta)$
(recall~\eqref{equ:phase-VT}) with respect to $\eta$ at $\eta=\xi$
(and $\norm{\xi}<1$); as before, we write
\[
  \sigma(\eta) = (f(\eta), \eta), \quad f(\eta) =
  \sqrt{1-\norm{\eta}^2}
\]
and use, with $\tilde \eta = \eta-\xi$:
\begin{equation}
  f(\xi+\tilde \eta) = f(\xi) - \frac{\pscal{\xi}{\tilde \eta}}{f(\xi)}
  - \frac{\norm{\tilde \eta}^2}{2f(\xi)} -
  \frac{\pscal{\xi}{\tilde \eta}^2}{2f(\xi)^3} + \O(\tilde \eta^3)\,.
  \label{equ:taylor_f}
\end{equation}
Hence
\begin{align}
  \label{equ:taylor_sigma}
  \sigma(\xi+\tilde \eta)
  & = \sigma(\xi) + \left(
    - \frac{\pscal{\xi}{\tilde \eta}}{f(\xi)}, \tilde \eta
    \right) -
    \left(\frac{\norm{\tilde \eta}^2}{2f(\xi)}
    + \frac{\pscal{\xi}{\tilde \eta}^2}{2f(\xi)^3} \right)e_0
    + g_\xi(\tilde \eta)\\
  & =:  \sigma(\xi)  + \quad \sigma_{1;\xi}(\tilde \eta) \quad
    + \quad \sigma_{2;\xi}(\tilde \eta)
    \qquad \qquad \quad + g_\xi(\tilde \eta)\,,
\end{align}
where $e_0:=(1,0,\dots,0)$ and $g_\xi = \O(\tilde \eta^3)$.
Using~\eqref{equ:phase-VT}, we write, accordingly:
\[
  \phi(y, \xi+\tilde \eta) = \phi(y, \xi) + \phi_{1;\xi}(y, \tilde \eta)
  + \phi_{2;\xi}(y, \tilde \eta) + G_\xi(y, \tilde \eta)
\]
and hence
\begin{equation}
  K_{V_G Q}(y,\xi) = \frac{ e^{\frac{i}{\h}\phi(y, \xi)}
  }{(2\pi\h)^d}\intint e^{\frac{i}{\h}\pscal{\tilde \eta}{\tilde x}}
  e^{\frac{i}{\h} (\phi_{1;\xi}(y, \tilde \eta) + \phi_{2;\xi}(y, \tilde \eta))} \tilde q(\xi+\tfrac{\tilde \eta}{2},\tilde x)
  \dd{\tilde x}\dd{\tilde \eta} \,.
  \label{equ:apparent-phase}
\end{equation}
with
\[
  \tilde q(\xi+\tfrac{\tilde \eta}{2},\tilde x) := e^{\frac{i}{\h}
    G_\xi(y, \tilde \eta)} q(\xi+\tfrac{\tilde \eta}{2},\tilde x) \,.
\]
Notice that $\phi_{1;\xi}$ is linear, $\phi_{2;\xi}$ is quadratic, and
$G_\xi(y, \tilde \eta) = \O(\tilde \eta^3)$.  This last estimate
implies that $\tilde q$ is ``slowly oscillating'' when $\tilde \eta$
is small, which enables the use of the stationary phase formula with
the apparent quadratic phase of~\eqref{equ:apparent-phase}.

The quadratic part of the phase is
\begin{align}
  \nonumber\mathcal{R}(\tilde x,\tilde \eta)
  & = \pscal{\tilde \eta}{\tilde x} - \phi_{2;\xi}(\tilde \eta) \\
  & = \pscal{\tilde \eta}{\tilde x}
    - \left(\frac{\norm{\tilde \eta}^2}{2f(\xi)}
    + \frac{\pscal{\xi}{\tilde \eta}^2}{2f(\xi)^3} \right)
    \pscal{y}{\pib({\trsp G}e_0)}\,.
\end{align}
Its Hessian matrix (of size $2d\times 2d$) has the form (in the
variables $(\tilde x, \tilde \eta)$)
\begin{equation}
  \mathcal{R} = \begin{pmatrix} 0 & \textup{Id} \\ \textup{Id} &
                                                                 -\frac{\alpha}{f} \left(\textup{Id} +\frac{1}{f^2} \xi\xi^t
                                                                 \right)
  \end{pmatrix}\label{equ:R}
\end{equation}
with $\alpha:= \alpha(y) = \pscal{y}{\pib({\trsp G}e_0)}$ and
$\xi\xi^t$ is the matrix $(\xi_i\xi_j)_{1\leq i,j\leq d}$. The
signature is zero and the determinant is 1. The inverse matrix is
\[
  \mathcal{R}^{-1} = \begin{pmatrix} \frac{\alpha}{f}
    \left(\textup{Id} +\frac{1}{f^2} \xi\xi^t \right) & \textup{Id} \\
    \textup{Id} & 0
  \end{pmatrix}
\]
We find, with
$D=\frac{1}{i}(\partial_{\tilde x}, \partial_{\tilde \eta})$,
\[
  \pscal{\mathcal{R}^{-1}D}{D} = -2\pscal{\partial_{\tilde x}}{\partial_{\tilde \eta}}
  - \frac{\alpha(y)}{f(\xi)}
  \left(\Delta_{\tilde x} +
    \frac{1}{f(\xi)^2}\pscal{\xi}{\partial_{\tilde x}}^2\right)\,.
\]
We apply again the quadratic stationary phase lemma:
\begin{align*}
  K_{V_G Q}(y,\xi) & \sim e^{\frac{i}{\h}\phi(y,\xi)} \sum_{k\geq
                    0}\frac{(i\h)^k}{k!}
                     \left(-\tfrac12\pscal{\mathcal{R}^{-1}D}{D}\right)^k \tilde q
                     (\xi+\tfrac{\tilde \eta}{2}, \tilde x)_{\tilde \eta = 0}
\end{align*}
and use the fact that the values of $\tilde q$ and its derivatives up
to order 2 on the critical set $\Lambda_{V_G Q}$ (where $\tilde\eta=0$)
are equal to those of $q$, to obtain
\begin{prop}\label{prop:KVQ} If $Q=\Op_\h^w(q_\h)$, with
  $q_\h=q_0 + \h q_1 + \O(\h^2)$, we have
  \begin{align}
    K_{V_G Q}(y,\xi)
    & = e^{\frac{i}{\h}\phi(y,\xi)}
      \left[  q_0(\xi,\tilde x) + \h q_1(x,\xi) \right.\nonumber\\
    & + \frac{i\h}2 \left(
      \pscal{\partial_{\tilde x}}{\partial_\xi}q_0(\xi,\tilde x) +
      \frac{\alpha\Delta_{\tilde x} q_0(\xi,\tilde x)}{f}
      + \frac{\alpha}{f^3} \pscal{\xi}{\partial_{\tilde x}}^2
      q_0(\xi,\tilde x) \right)
      \nonumber\\
    &  \left.  + \O(\h^2)\right]\,,\label{equ:KVTQ}
  \end{align}
  where $f = f(\xi) = \sqrt{1-\norm{\xi}^2}$,
  $\alpha =\alpha(y) = \pscal{y}{\pib({\trsp G}e_0)}$, and
  $\tilde x = -\partial_\xi\phi(y,\xi) = - {\trsp {\dd g}}(\xi)\cdot
  y$.
\end{prop}

Propositions~\ref{prop:KPV} and~\ref{prop:KVQ} offer a way to
computing the actions of phase space filters on the transformed
hologram directly from the initial hologram $\psi_0$; they are also
instrumental in proving the Egorov theorem mentioned earlier, as we
shall see in the next section.

\subsection{Conjugation by $U_G$}
Let us now turn to the Egorov property. Assume that $Q=\Op_\h^w(q_\h)$
and $P=\Op_\h^w(p_h)$ are pseudodifferential operators related by the
(microlocal) equation
\[
  PV_G = V_G Q\,.
\]
Let $(x,\xi)\in\RM^{2d}$ and assume that
$\sigma(\xi)\not\in G\mathcal{P}_0$ (see
Proposition~\ref{prop:hessian}). From Proposition~\ref{prop:KPV} and
Proposition~\ref{prop:KVQ} we get, by equating terms of order zero in
$\h$ in $K_{PV_G}(x,\xi) = K_{V_G Q}(x,\xi)$:
\begin{equation}
  p_0(x,\partial_x \phi(x,\xi)) = q_0(\xi,-\partial_{\xi}\phi(x, \xi))
  \label{equ:p0-q0}
\end{equation}
(indeed, the term $e^{\frac{i}{\h}\pscal{x}{\tilde\xi}}$
in~\eqref{equ:KPVT} is equal to $e^{\frac{i}{\h}\phi(x,\eta)}$). In
other words, $q_0 = p_0\circ \kappa_{V_G}$, where $\kappa_{V_G}$ is
the canonical transformation~\eqref{equ:canonical-transform-V}. We
have simply recovered the usual second statement of the Egorov
theorem.

Our new result is the computation of the term of order $\h$. We obtain
the implicit equation
\begin{equation}\label{equ:p1-0}
  p_1 - q_1 = \frac{i}{2} \left(\pscal{\partial_{\tilde \xi}}{\partial_x}p_0
    + \pscal{\partial_{\tilde x}}{\partial_\xi}q_0 +
    \frac{\alpha(x)}{f(\xi)}\Delta_{\tilde x} q_0 +
    \frac{\alpha(x)}{f(\xi)^3} \pscal{\xi}{\partial_{\tilde x}}^2q_0\right)
\end{equation}
where $p_0,p_1$ are evaluated at
$(x,\tilde\xi = \partial_x \phi(x,\xi))$ and $q_0,q_1$ are evaluated
at $(\xi,\tilde x = -\partial_{\xi}\phi(x, \xi))$.

We may now come back to the original operator
$U_G= V_G\mathcal{F}_\h$. In order to obtain a more pleasant
formulation, we introduce a \emph{microlocal left inverse} $U_G^{-1}$
of $U_G$.  In terms of wavefronts, $U_G$ transports wave functions
microlocalized near $(x,\xi)$ to wave functions microlocalized near
$\kappa_{U_G}(x,\xi)$. Since $\kappa_{U_G}$ might not be injective
(see Theorem~\ref{theo:canonical}), its inverse is multivalued. By
definition, we call $U_G^{-1}$ the FIO that satisfies
$U_G^{-1} U_G \sim \textup{Id}$ near $(x,\xi)$. We shall also denote
by $g^{-1}(\tilde \xi)$ a left inverse of $g$ defined near $g(\xi)$.

\begin{theo}\label{theo:egorov}
  Let $P$ be a semiclassical pseudodifferential operator with Weyl
  symbol $p=p_0$ independent of $\h$. Then, in a phase space region
  where $\norm{\xi}\leq 1-\epsilon$, $\epsilon>0$, and
  $\sigma(\xi)\not\in G\mathcal{P}_0$, the Weyl symbol of
  $R=U_G^{-1} P U_G$ is $r_0(x,\xi) + hr_1(x,\xi)+ \O(\h^2)$ with
  \begin{equation}
    r_0 = p_0\circ \kappa_{U_G}
    \label{equ:p0}
  \end{equation}
  and
  \begin{equation}
    r_1= \frac{i}{2J}\{J,r_0\} = \left[
      \frac{i}{2\tilde J}\{\tilde J,p_0\}
    \right] \circ \kappa_{U_G}
    \label{equ:p1}
  \end{equation}
  where
  \[
    J(\xi) := \det (\dd g(\xi))^{-1}\,, \quad \tilde{J}(\tilde\xi) =
    J(g^{-1}(\tilde \xi)) = \det (\dd g^{-1}(\tilde\xi))
  \]
  and $\{\cdot,\cdot\}$ is the Poisson bracket.
\end{theo}
\begin{demo}
  Given any pseudodifferential operator $R$, we have
  \[
    U_G R U_G^{-1} = V_G (\mathcal{F}_\h R \mathcal{F}_\h^{-1})
    V_G^{-1} = V_G Q V_G^{-1}
  \]
  with $Q:=\mathcal{F}_\h R \mathcal{F}_\h^{-1}$. Since
  $\mathcal{F}_\h$ is a metaplectic operator, the Weyl symbols of $Q$
  and $R$ are related by a linear change of variables, namely
  \[
    q_\h(\xi,-x) = r_\h(x, \xi)\,.
  \]
  Taking $R=U_G^{-1} P U_G$, we have $P=V_G Q V_G^{-1}$,
  hence~\eqref{equ:p0-q0} becomes
  $r_0(\partial_\xi\phi, \xi) = p_0(x,\partial_x\phi)$, which
  gives~\eqref{equ:p0}. Moreover, in view of~\eqref{equ:p1-0}, we have
  \begin{equation}
    - r_1(\partial_\xi\phi, \xi) = \frac{i}{2}\left[
      \pscal{\partial_{\xi}}{\partial_x}p_0 -
      \pscal{\partial_{\xi}}{\partial_x}r_0 +
      \frac{\alpha(x)}{f(\xi)}\Delta_{x} r_0 +
      \frac{\alpha(x)}{f(\xi)^3} \pscal{\xi}{\partial_{x}}^2r_0\right]
    \,.
    \label{equ:p1-1}
  \end{equation}

  Let us now compute the first term,
  $\pscal{\partial_{\xi}}{\partial_x}p_0 -
  \pscal{\partial_{\xi}}{\partial_x}r_0$. When $T=\textup{Id}$, this
  term vanishes (and hence $p_1=0$, naturally, since $\alpha=0$). In
  general, it can be computed in terms of $r_0$, as follows.

  Taking the derivative of~\eqref{equ:p0} with respect to $x$, we
  obtain
  $\partial_x p_0 + \partial_\xi p_0 \cdot \partial^2_x\phi =
  \partial_x r_0 \cdot \partial_{x}\partial_{\xi}\phi$. Remember
  from~\eqref{equ:phase-VT} that $\phi$ is linear in $x$:
  \[
    \phi(x, \xi) = \pscal{x}{\pib {\trsp G}\sigma(\xi)}_{\RM^d},
  \]
  and hence $\partial^2_x\phi=0$, which gives
  \begin{equation}
    \label{equ:dxp}
    \partial_x p_0 = \partial_x r_0 \cdot \partial_{x}\partial_{\xi}\phi\,.
  \end{equation}
  As usual, when $f=f(x,\xi)$ is a function on $\RM^{2d}$, we denote
  by $\partial_x f$ the partial differential with respect to
  $x\in \RM^d$, which is a linear map on the tangent space to the $x$
  variable (and similarly for $\partial_{\xi}f$). The term
  $\partial_{x}\partial_{\xi}\phi$ is a linear endomorphism of the
  tangent space $\Tg_x\RM^d$. Thus,~\eqref{equ:dxp} is an equality in
  the space of linear forms (covectors), \emph{i.e.} for any
  $u\in \Tg_x\RM^d$:
  \[
    \partial_x p_0 (u) = \partial_x r_0
    \big(\partial_{x}[\partial_{\xi}\phi](u)\big)\,.
  \]
  (In this whole computation, for notation simplicity, $p_0$ and its
  derivatives are evaluated at $(x,\partial_x\phi)$ and $r_0$ and its
  derivatives are evaluated at $(\partial_\xi\phi, \xi)$, just like
  in~\eqref{equ:p0}, while $\phi$ and its derivatives are evaluated at
  $(x,\xi)$.) Taking now the derivative of~\eqref{equ:dxp} with
  respect to $\xi$, we get, for any $v\in \Tg_{\xi}\RM^d$, the
  following equality of linear forms:
  \begin{gather}
    \label{equ:dxdxip_phi}
    \partial_{\xi}[\partial_x p_0] \cdot
    \partial_{\xi}[\partial_x\phi](v) = \left(\partial^2_x r_0 \cdot
      \partial^2_\xi\phi(v) + \partial_{\xi}[\partial_x
      r_0](v)\right)\cdot \partial_{x}\partial_{\xi}\phi \\\nonumber +~
    \partial_x r_0\cdot
    \big(\partial_{\xi}[\partial_x\partial_{\xi}\phi](v)\big)\,.
  \end{gather}
  If we denote by $A$ the endomorphism
  $A:=\partial_{x}\partial_{\xi}\phi$ of $\Tg_x\RM^d$, we have
  \[
    \partial_{\xi}\partial_x p_0 \cdot A^t = A^t \cdot (\partial^2_x
    r_0 \cdot \partial^2_\xi\phi + \partial_{\xi}\partial_x r_0) + B
  \]
  where we denote by $B=B_{x,\xi}$ the endomorphism of
  $\Tg_{\xi}\RM^d = (\Tg_x\RM^d)^*$ defined by
  \begin{equation}
    B(v) = \partial_x r_0\cdot \big(\partial_{\xi} A(v)\big)\,.
    \label{equ:B}
  \end{equation}
  Recall that the mixed hessian matrix $A$ was computed in
  Proposition~\ref{prop:hessian}; $A={\trsp {\dd g}(\xi)}$, so it
  depends only on $\xi$, and is invertible under the conditions
  mentioned there. Multiplying~\eqref{equ:dxdxip_phi} on the right by
  the $(A^t)^{-1}$, we get
  \begin{equation}
    \label{equ:dxdxip}
    \partial_{\xi}\partial_x p_0 = A^t \cdot (\partial^2_x r_0
    \cdot \partial^2_\xi\phi + \partial_{\xi}\partial_x r_0)\cdot (A^t)^{-1}
    + B\cdot (A^t)^{-1}\,.
  \end{equation}
  We now compute
  $\partial^2_\xi\phi = \pscal{x}{\pib {\trsp
      G}\partial_\xi^2\sigma(\xi)}$.  The hessian
  $\partial_\xi^2\sigma(\xi)$ was computed, as a quadratic form,
  in~\eqref{equ:taylor_sigma} (this is the term $\sigma_{2;\xi}$); and
  then $\partial^2_\xi\phi$ is the lower right term of~\eqref{equ:R},
  \emph{i.e.}
  \begin{equation}
    \partial^2_{\xi} \phi = -\frac{\alpha}{f} \left(\textup{Id}
      +\frac{1}{f^2} \xi\xi^t\right)\,.
    \label{equ:d2xi_phi}
  \end{equation}
  One could also write
  \[
    \phi(x,\xi) = \pscal{\pib T X}{\xi} + f(\xi)\pscal{TX}{e_3} =
    \pscal{\pib T X}{\xi} + f(\xi)\alpha(x)\,
  \]
  where $X:=(x,0)\in\RM^{1+d}$, and hence
  \[
    \partial^2_\xi\phi = \alpha(x) \dd{}^2f.
  \]
  Using~\eqref{equ:taylor_f}, we obtain $\dd f = -\frac{1}{f}\xi^t$
  and
  ${\dd{}}^2f = -\frac{1}{f}(\textup{Id} + \frac{1}{f^2}\xi\xi^t)$,
  which gives~\eqref{equ:d2xi_phi} again.

  Let us now take the trace of the equality~\eqref{equ:dxdxip}, viewed
  as $2\times 2$ matrices:
  \[
    \tr \partial_{\xi}\partial_x p_0 = \tr (\partial^2_x r_0 \cdot
    \partial^2_\xi\phi) + \tr \partial_{\xi}\partial_x r_0 + \tr
    (B\cdot (A^t)^{-1})\,.
  \]
  We have
  \[
    \tr \partial_{\xi}\partial_x p_0 = \sum_{i=1}^2
    \partial_{x_i}\partial_{\xi_i} p_0 =
    \pscal{\partial_{\xi}}{\partial_x}p_0, \qquad\tr
    \partial_{\xi}\partial_x r_0 =
    \pscal{\partial_{\xi}}{\partial_x}r_0\,.
  \]
  And
  \[
    \partial^2_x r_0 \cdot \partial^2_\xi\phi =
    -\frac{\alpha}{f}\partial^2_x r_0 -
    \frac{\alpha}{f^3}\left(\partial^2_x r_0 \cdot \xi\xi^t\right)
  \]
  which gives
  \[
    \tr \partial^2_x r_0 \cdot \partial^2_\xi\phi = -
    \frac{\alpha}{f}\Delta_x r_0 - \frac{\alpha}{f^3}
    \pscal{\xi}{\partial_x}^2r_0\,.
  \]
  Hence,
  \[
    \pscal{\partial_{\xi}}{\partial_x}p_0 -
    \pscal{\partial_{\xi}}{\partial_x}r_0 = - \frac{\alpha}{f}\Delta_x
    r_0 - \frac{\alpha}{f^3} \pscal{\xi}{\partial_x}^2r_0 + \tr \left(
      B\cdot (A^t)^{-1}\right)\,.
  \]
  Thus, in Formula~\eqref{equ:p1-1}, several cancellations occur, we
  simply obtain
  \begin{equation}
    - r_1(\partial_\xi\phi, \xi) = \frac{i}{2}\left[
      \tr \left( B\cdot (A^t)^{-1}\right)\right]\,.\label{equ:p1-2-tmp}
  \end{equation}
  It remains to compute the trace:
  $ \tr \left( B\cdot (A^t)^{-1}\right)$. Using the commutation of the
  derivatives with respect to $\xi$, the $2\times2\times 2$-tensor
  $\partial_\xi A = \partial_\xi \partial_x \partial_\xi \phi$
  satisfies:
  \[
    w\cdot (\partial_\xi A) (v) = v \cdot (\partial_\xi A)(w), \quad
    \forall v,w \in \Tg_\xi\RM^d\,.
  \]
  Hence $B(v) = v \cdot (\partial_\xi A)(\partial_x r_0) $, which
  means that $B=[(\partial_\xi A)(\partial_x r_0)]^t$. Therefore
  $\tr B\cdot (A^t)^{-1} = \tr A^{-1}\cdot (\partial_\xi A)(\partial_x
  r_0)$.  For any covector $w=(w_1,w_2)$ we have
  \[
    A^{-1}\cdot (\partial_\xi A)(w) = \sum_j w_j A^{-1}\cdot
    \partial_{\xi_j}A = \sum_j w_j \partial_{\xi_j}(\log A)\,.
  \]
  Hence
  \[
    \tr B\cdot (A^t)^{-1} = \sum_j (\partial_{x_j}r_0)
    \partial_{\xi_j}(\tr (\log A)) = \sum_j (\partial_{x_j}r_0)
    \partial_{\xi_j} (\log (\det A)).
  \]
  Let us denote by $J=J(\xi)$ the Jacobian determinant
  $J=\det A^{-1} = (\det \dd g (\xi))^{-1}$. (Recall that
  $ \dd g (\xi)$ is invertible by assumption due to
  Proposition~\ref{prop:hessian}.) We have
  \[
    \tr B\cdot (A^t)^{-1} = - \frac{1}{J} \sum_j (\partial_{x_j}r_0)
    \partial_{\xi_j} J = -\frac{1}{J} \pscal{\partial_x
      r_0}{\partial_\xi J} = - \frac{1}{J}\{J,r_0\}
  \]
  (since $J$ only depends on $\xi$,
  $\{J,r_0\} = \pscal{\partial_x r_0}{\partial_\xi J}\,$.)  Now,
  using~\eqref{equ:p0} and the fact that $\kappa_{U_G}$ is symplectic,
  \[
    \{J,r_0\} = \{J,p_0\circ \kappa_{U_G}\} = \{J\circ
    \kappa_{U_G}^{-1}, p_0\} \circ \kappa_{U_G}
  \]
  and
  $J\circ \kappa_{U_G}^{-1}(\tilde x,\tilde \xi) = J\circ
  g^{-1}(\tilde \xi)$ which, together with~\eqref{equ:p1-2-tmp},
  finally proves~\eqref{equ:p1}.
\end{demo}

\begin{rema}
  Our proof of Theorem~\ref{theo:egorov} proceeds by direct
  computation; we believe that it is worth presenting here because we
  were not able to find similar calculations in the
  literature. However, it may not shed light on the various
  cancellations which give rise to the simple
  formula~\eqref{equ:p1}. In Section~\ref{sec:lack-unitarity} below,
  we present an indirect proof based on the fact that the Weyl symbol
  of a symmetric operator must be real valued, see
  Remark~\eqref{rema:real-valued}. It is also probable that a more
  conceptual proof could be derived from the microlocal
  formula~\eqref{equ:factorized-microlocal}, using the fact that
  changes of variables preserve the subprincipal symbol of
  pseudodifferential operators when acting on half-densities.
\end{rema}

\section{Lack of unitarity}\label{sec:lack-unitarity}

When $G=\textup{Id}$, we have
$V_G = \mathcal{F}_\h^{-1} = \frac{1}{(2\pi\h)^d}\mathcal{F}_\h^*$,
and hence $(2\pi\h)^{d/2} V_G$ is unitary on $L^2(\RM^d)$. When
$\mathcal{G}=\tau_\gamma$ is a translation,
$(2\pi\h)^{d/2} V_{\mathcal{G}}$ is also unitary, due
to~\eqref{equ:V-gamma}. Hence, thanks to
Lemma~\ref{lemm:decomposition}, for a general affine transformation
$\mathcal{G}$, the unitarity of $V_\mathcal{G}$ (or, equivalently,
$U_\mathcal{G}$) reduces to the unitarity of $V_G$, where $G$ is the
linear part of $\mathcal{G}$.

For a general $G$, the operators $(2\pi\h)^d V_G^* V_G$ and
$(2\pi\h)^d V_G V_G^*$ are not the identity. Recall that the integral
kernel of $V_G$ is
\[
  K_{V_G}(y, \eta) = \frac{ e^{\frac{i}{\h}\phi(y, \eta)}}{(2\pi\h)^d}
\]
and hence the integral kernel of $V_G^*$ is
\[
  K_{V_G^*}(\eta, x) = \overline{K_{V_G}(x,\eta)} = \frac{
    e^{-\frac{i}{\h}\phi(x, \eta)}}{(2\pi\h)^d}\,.
\]
Therefore the integral kernel of the composition $V_G V_G^*$ is
\[
  K_{V_G V_G^*}(y,x) = \int K_{V_G}(y, \eta)K_{V_G^*}(\eta, x)
  \dd{\eta} = \frac{1}{(2\pi\h)^{2d}}\int e^{\frac{i}{\h}(\phi(y,
    \eta) - \phi(x, \eta))} \dd{\eta}\,.
\]
We have
\[
  \phi(y, \eta) - \phi(x, \eta) = \pscal{y-x}{\pib {\trsp G}
    \sigma(\eta)}\,.
\]
In general, this phase may be degenerate; however, if we restrict to
the set of $\eta$ such that $\sigma(\eta)\not\in G\mathcal{P}_0$, then
Proposition~\ref{prop:hessian} implies that the phase is
non-degenerate, and the corresponding Lagrangian submanifold is
included in the diagonal. In this case, the composition is
microlocally an FIO associated with the identity canonical
transformation, hence a pseudodifferential operator.  Let us compute
it explicitly.

We consider the ``change of variables''
$\tilde\eta = g(\eta) := \pib {\trsp G} \sigma(\eta)$; as we saw in
Section~\ref{sec:study-v_G}, this map is in general
non-injective. More precisely, we use the
decomposition~\eqref{equ:decomposition_D}; on each domain $D_j$,
$j=\pm$, we have a smooth left inverse for $g : D_j \to g(D_j)$, which
we denote by $g^{-1}_j$:
\[
  g_j^{-1}(g(\eta)) = \eta \quad \text{for } \eta\in D_j = \pib {\trsp
    G}^{-1}\mathcal{E}_j\,.
\]

Instead of $V_{G}$, we consider $\tilde V := V_{G} \chi$, where
$\chi=\chi(\eta)$ is a smooth cut-off function compactly supported in
the unit disc $\norm{\eta} < 1$. Then
\begin{align*}
  K_{\tilde V \tilde V^*} (y,x)
  & =  \frac{1}{(2\pi\h)^{2d}}\int_{\norm{\eta} < 1} e^{\frac{i}{\h}(\phi(y,
    \eta) - \phi(x, \eta))} \chi^2(\eta) \dd{\eta} \\
  & =  \frac{1}{(2\pi\h)^{2d}} \left[ \int_{g(D_-)}\!\!\!\!\!\!\!e^{\frac{i}{\h}\pscal{y-x}{\tilde{\eta}}}
    J_-(\tilde \eta)  \rho_-(\tilde \eta)
    \dd{\tilde{\eta}} +
    \int_{g(D_+)}\!\!\!\!\!\!\!e^{\frac{i}{\h}\pscal{y-x}{\tilde{\eta}}}
    J_+(\tilde \eta) \rho_+(\tilde \eta)
    \dd{\tilde{\eta}} \right]
\end{align*}
where $J_j(\tilde\eta)$ is the Jacobian determinant of $g^{-1}_j$ and
$ \rho_j(\tilde \eta) = \chi^2(g^{-1}_j(\tilde\eta))$.  Hence
\begin{align*}
  K_{\tilde V \tilde V^*}  (y,x)
  & =\frac{1}{(2\pi\h)^{2d}}
    \int_{\RM^d} e^{\frac{i}{\h}\pscal{y-x}{\tilde{\eta}}}
    ( J_+(\tilde \eta) + J_-(\tilde \eta)  )
    \dd{\tilde{\eta}}\\
  & = \frac{1}{(2\pi\h)^d} \mathcal{F}_\h^{-1}
    \tilde J  \mathcal{F}_\h\,, \,,
\end{align*}
where $\tilde J = J_+ + J_-$ and
\[
  J_j(\tilde \eta) := J_j(\tilde\eta)\chi^2(g^{-1}_j(\tilde\eta))
  \mathbf{1}_{g(D_j)}\,.
\]
For instance, if $\tilde\eta\in g(D_+)\cap g(D_-)$, then there exist
$\eta_j\in D_j$ (hence $\eta_+\neq \eta_-$), such that
$g(\eta_+) = g(\eta_-) = \tilde\eta$. In this case
$\tilde J(\tilde \eta)$ really has two contributions from two
different frequencies.

In general, acting on functions whose frequency variable is localised
in $g(D_+)\cup g(D_-)$ we obtain the microlocal equality
\[
  V_G V_G^* = \frac{1}{(2\pi\h)^d} \mathcal{F}_\h^{-1} \tilde J
  \mathcal{F}_\h\,.
\]

Consider now the operator $V_G^* V_G$; its integral kernel is
\[
  K_{V_G^* V_G}(\xi,\eta) = \int K_{V_G^*}(\xi, y)K_{V_G}(y, \eta) \dd
  y = \frac{1}{(2\pi\h)^{2d}}\int e^{\frac{i}{\h}(\phi(y, \eta) -
    \phi(y, \xi))} \dd y\,.
\]
Let us consider the phase
$\varphi(\xi,\eta,y):= \phi(y, \eta) - \phi(y, \xi)$. We have
\[
  \varphi(\xi,\eta,y) = \pscal{GY}{\sigma(\eta) -
    \sigma(\xi)}_{\RM^{1+d}}\,,
\]
where we denote $Y:=(0,y) = (0,y_1,\dots,y_d)\in\RM^{1+d}$. We have
\[
  \sigma(\eta) - \sigma(\xi) = (f(\eta) - f(\xi), \eta - \xi)\,.
\]
On the other hand,
\begin{align}
  \nonumber(f(\eta) - f(\xi))(f(\eta) + f(\xi))
  & = f(\eta)^2 - f(\xi)^2\\\nonumber
  & = \norm{\xi}^2 - \norm{\eta}^2\\
  & = \pscal{\xi-\eta}{\xi+\eta}\,.\label{equ:f(eta)-f(xi)}
\end{align}
Therefore,
\[
  \varphi(\xi,\eta,y) = \pscal{GY}{(0,\eta)-(0,\xi) +
    \frac{1}{f(\xi)+f(\eta)}\pscal{\xi-\eta}{\xi+\eta}e_0}_{\RM^{1+d}}
\]
which can be written
\[
  \varphi(\xi,\eta,y) = \pscal{GY}{B \cdot (\eta-\xi)}_{\RM^{1+d}}
\]
with
\[
  B = B(\xi,\eta) :=
  \begin{pmatrix}
    \frac{-(\xi+\eta)^t}{f(\xi) + f(\eta)}\\
    \textup{Id}
  \end{pmatrix}\, : \RM^d \to \RM^{1+d}
\]
Hence
\[
  \varphi(\xi,\eta,y) = \pscal{{\trsp B} GY}{\eta-\xi} \,.
\]
For fixed $(\xi,\eta)$, consider the change of variables
$\tilde y = g_{\xi,\eta}( y) := {\trsp B} GY$; we have
\[
  \int e^{\frac{i}{\h}\varphi(\xi,\eta,y)} \dd y = \int
  e^{\frac{i}{\h}\pscal{\tilde y}{\eta-\xi}}J_*(\xi,\eta,\tilde y)
  \dd{\tilde y}
\]
where $J_*(\xi,\eta,\tilde y)$ is the Jacobian determinant of
$g_{\xi,\eta}^{-1}$. Since $g_{\xi,\eta}$ is linear in $y$, the
Jacobian does not depend on $\tilde y$, and we can write
\[
  \int e^{\frac{i}{\h}\varphi(\xi,\eta,y)} \dd y = J_*(\xi,\eta) \int
  e^{\frac{i}{\h}\pscal{\tilde y}{\eta-\xi}} \dd{\tilde y} =
  (2\pi\h)^d J_*(\xi,\eta) \delta_{(\eta-\xi)}
\]
We finally find that the operator $V_G^* V_G$ is simply the
multiplication operator
$u(\xi) \mapsto \frac{1}{(2\pi\h)^d}J_*(\xi,\xi) u(\xi)$, and
\[
  J_*(\xi,\xi) = \frac{1}{\det ({\trsp B(\xi,\xi)} G \pib^{*})}\,.
\]
Comparing~\eqref{equ:f(eta)-f(xi)} with the Taylor expansion of $f$ at
$\eta=\xi$, we see that $\dd f(\xi) = \frac{-\xi^t}{f(\xi)}$ and hence
$\dd \sigma (\xi) = B(\xi,\xi)$ (this was actually already computed
in~\eqref{equ:taylor_sigma}). This proves that
$J_*(\xi,\xi) = J(\xi)$; indeed:
\begin{align*}
  \frac{1}{J(\xi)}
  &= \det (\dd{}_\xi (\pib {\trsp G} \sigma)) =
    \det (\pib {\trsp G} \dd{}_\xi \sigma)\\
  &=  \det (\pib {\trsp G} B(\xi,\xi)) = \det ({\trsp B(\xi,\xi)} G \pib^*)\\
  &= \frac{1}{J_*(\xi,\xi)}\,.
\end{align*}

Let us summarise this in the following proposition.
\begin{prop}
  Let $J(\xi) := (\det \dd g(\xi))^{-1}$.  We have, microlocally near
  any $(x,\xi)$ such that $\norm{\xi}<1$ and
  $\sigma(\xi)\not\in G\mathcal{P}_0$,
  \[
    V_G V_G^* = \frac{1}{(2\pi\h)^d} \mathcal{F}_\h^{-1} \tilde J
    \mathcal{F}_\h\,,
  \]
  where $\tilde J$ is the multiplication operator by the function
  \[
    \tilde J(\xi) = \sum_{\eta \in g^{-1}(g(\xi))} J(\eta)\,.
  \]
  And, microlocally near any $(\xi,x)$ such that $\norm{\xi}<1$ and
  $\sigma(\xi)\not\in G\mathcal{P}_0$,
  \[
    V_G^* V_G = \frac{1}{(2\pi\h)^d} J\,,
  \]
  where $J$ is the multiplication operator by the function
  $J= (\det \dd g(\xi))^{-1}$.
\end{prop}
\begin{coro}\label{coro:lack-unitarity-U}
  We have, microlocally near any $(x,\xi)$ such that $\norm{\xi}<1$
  and $\sigma(\xi)\not\in G\mathcal{P}_0$,
  \[
    U_G U_G^* = \mathcal{F}_\h^{-1} \tilde J \mathcal{F}_\h
  \]
  and
  \[
    U_G^* U_G = \mathcal{F}_\h^{-1} J \mathcal{F}_\h\,.
  \]
\end{coro}

We may now answer the question raised in the beginning of
Section~\ref{sec:prec-egor-theor}.
\begin{theo}\label{theo:egorov-adjoint}
  Let $G\in \textup{GL}_{1+d}(\RM)$ and $\gamma\in\RM^{1+d}$. Let
  $\mathcal{G}:=\tau_\gamma\circ G$ be the corresponding affine
  transformation.  Let $P$ be a semiclassical pseudodifferential
  operator with Weyl symbol $p=p_0$ independent of $\h$. Then, in a
  phase space region where $\norm{\xi}\leq 1-\epsilon$, $\epsilon>0$,
  and $\sigma(\xi)\not\in G\mathcal{P}_0$, the Weyl symbol of
  $T=U_{\mathcal{G}}^{*} P U_\mathcal{G}$ is $t_0+\O(\h^2)$ (that is,
  the subprincipal term $\h t_1$ vanishes), with
  \begin{equation}
    t_0(x,\xi) = J(\xi)p_0\circ \kappa_{U_\mathcal{G}}(x,\xi)
    \label{equ:s0}
  \end{equation}
\end{theo}
\begin{demo}
  Thanks to Lemma~\ref{lemm:decomposition} (item~\ref{item:gamma-G}),
  it is enough to consider separately the cases $\mathcal{G}=G$ and
  $\mathcal{G}:=\tau_\gamma$.

  \paragraph{Case $\mathcal{G}=G$.} Let $R=U_G^{-1} P U_G$.  We know
  from Theorem~\ref{theo:egorov}, that the Weyl symbol of $R$ is
  $r_0 + i\h \tilde r_1 + \O(\h^2)$, for some smooth function
  $\tilde r_1$, and $r_0=p_0\circ \kappa_{U_G}$. Writing
  $T = U_G^{*} U_G R$ and applying
  Corollary~\ref{coro:lack-unitarity-U}, we have $T = \hat J R$, where
  we denote by $\hat J$ the Fourier multiplier
  $\mathcal{F}_\h^{-1} J \mathcal{F}_\h$, we may derive the Weyl
  symbol $t_\h$ of $T$ by the composition formula for
  pseudodifferential operators (Moyal's formula\footnote{due to
    Groenewold~\cite{groenewold}}, see~\cite[Theorem
  4.12]{zworski-book-12}): if $A=\Op_\h^w(a_\h)$ and
  $B=\Op_\h^w(b_\h)$, and $a_\h$ and $b_\h$ admit asymptotic
  expansions in powers of $\h$, then
  \[
    \sigma_W(A\circ B) = a_\h b_\h + \frac{\h}{2i}\{a_\h,b_\h\} +
    \O(\h^2)\,.
  \]
  In our situation, this gives
  \[
    t_\h(x,\xi) = J(\xi) r_0(x,\xi) + i\h J(\xi)\tilde r_1(x,\xi) +
    \frac{\h}{2i}\{J,r_0\} + \O(\h^2)\,.
  \]
  Assume that $P$ is symmetric, so that its symbol is real
  valued. Then it follows from Theorem~\ref{theo:egorov} that $r_0$
  and $\tilde r_1$ are real valued. Since $T$ is then also symmetric,
  $t_\h$ must be real valued as well. This implies
  \begin{equation}
    J(\xi)\tilde r_1(x,\xi) = \frac{1}{2}\{J,r_0\}\label{equ:real-valued}
  \end{equation}
  and hence
  \[
    t_\h(x,\xi) = J(\xi) r_0(x,\xi) + \O(\h^2)\,,
  \]
  which proves the theorem.
  \begin{rema}\label{rema:real-valued}
    It is easy to obtain $r_\h=r_0 + i\h \tilde r_1 + \O(\h^2)$, for
    some real valued function $\tilde r_1$, from
    Propositions~\ref{prop:KPV} and~\ref{prop:KVQ}; namely,
    $\tilde r_1 = -i r_1$ with the formula for $r_1$ given
    by~\eqref{equ:p1-1}. Hence, the above argument shows
    that~\eqref{equ:real-valued} holds, and therefore
    \[
      \tilde r_1(x,\xi) = \frac{1}{2J}\{J,r_0\}
    \]
    which directly recovers~\eqref{equ:p1}, without further
    computation.
  \end{rema}

  \noindent \paragraph{ Case $\mathcal{G}=\tau_\gamma$.} This case can
  be treated directly by a stationary phase argument, but let us
  consider here a more general route, based upon the following lemma:
  \begin{lemm}\label{lemm:conjug-phase}
    Let $f=f(x)$ be a smooth real function and $A=\Op_\h^w(a)$ be a
    pseudodifferential operator. Then
    \[
      A_f := e^{-\frac{i}{\h}f}\circ A \circ e^{\frac{i}{\h}f}
    \]
    is a pseudodifferential operator with Weyl symbol
    \[
      a_f(x,\xi) := a(x,\xi+\dd f(x)) + \O(\h^2)\,.
    \]
  \end{lemm}
  \begin{demo}[of the lemma]
    This lemma is well known to specialists; obtaining the remainder
    $\O(\h)$ is standard and holds for any quantization; the vanishing
    of the subprincipal term is due to Weyl's quantization; we present
    an explicit argument (which in principle can be used for computing
    the expansion at any order) for the convenience of the
    reader. First, it is easy to see that, for any differential
    operator $A=\sum_{k=0}^m b_k(x)(\frac{\h}{i}\partial_x)^k$,
    \[
      e^{-\frac{i}{\h}f}\circ A \circ e^{\frac{i}{\h}f} = \sum_{k=0}^m
      b_k(x)\left(\frac{\h}{i}\partial_x + \nabla f(x)\right)^k\,.
    \]

    Using inductively Moyal's formula, we check that the Weyl symbol
    of $\left(\frac{\h}{i}\partial_x + \nabla f(x)\right)^k$ is
    exactly $(\xi+\dd f)^k$.  Hence, applying the Moyal formula again,
    for the products
    $b_k(x)\left(\frac{\h}{i}\partial_x + \nabla f(x)\right)^k =
    \Op_\h^w(b_k)\circ \Op_\h^w((\xi+\dd f)^k)$, we obtain
    \[
      a_f(x,\xi) = a_1(x,\xi+\dd f(x)) -
      \frac{\h}{2i}\pscal{\partial_x}{\partial_\xi} a_1 (x,\xi+\dd
      f(x)) + \O(\h^2)\,,
    \]
    where $a_1(x,\xi) := \sum_{k=0}^m b_k(x)\xi^k$ is the left-symbol
    of $A$. By density, and localisation in phase space, this remains
    true for any symbol $a_1(x,\xi)$ in a good symbol class. Finally,
    applying the formula that relates the Weyl symbol
    $a_{\frac{1}{2}}$ to the left symbol $a_1$:
    \[
      a_{\frac{1}{2}}(x,\xi) =
      \exp(\frac{i\h}{2}\pscal{\partial_x}{\partial_\xi}) a_1(x,\xi) =
      a_1(x,\xi) - \frac{\h}{2i}\pscal{\partial_x}{\partial_\xi}
      a_1(x,\xi)\,,
    \]
    we obtain the required formula for $a_f$.
  \end{demo}
  \paragraph{End of the proof of Theorem~\ref{theo:egorov-adjoint}.}~\\
  Recall from~\eqref{eq:U-gamma} that
  $ U_\gamma = \mathcal{F}_\h^{-1}
  e^{\frac{i}{\h}\pscal{\gamma}{\sigma(\cdot)}}\mathcal{F}_\h$. Hence
  \[
    U_\gamma^* P U_\gamma =
    \mathcal{F}_\h^{-1}e^{-\frac{i}{\h}\pscal{\gamma}{\sigma(\cdot)}}
    \mathcal{F}_\h P \mathcal{F}_\h^{-1}
    e^{\frac{i}{\h}\pscal{\gamma}{\sigma(\cdot)}}\mathcal{F}_\h\,.
  \]
  The Weyl symbol of $ \mathcal{F}_\h P \mathcal{F}_\h^{-1}$ is
  $p(-\xi,x)$.  By Lemma~\ref{lemm:conjug-phase}, the Weyl symbol of
  $e^{-\frac{i}{\h}\pscal{\gamma}{\sigma(\cdot)}} \mathcal{F}_\h P
  \mathcal{F}_\h^{-1} e^{\frac{i}{\h}\pscal{\gamma}{\sigma(\cdot)}}$
  is hence $p(-\xi-\dd S(x),x) + \O(\h^2)$ with
  $S:=\frac{i}{\h}\pscal{\gamma}{\sigma(\cdot)}$. Finally, the Weyl
  symbol of $ U_\gamma^* P U_\gamma$ is
  $p(x-\dd S(\xi),\xi) + \O(\h^2)$, and $p(x-\dd S(\xi),\xi)$ which is
  precisely $p\circ \kappa_{U_\gamma}(x,\xi)$ according to
  Theorem~\ref{theo:canonical}.
\end{demo}

\paragraph{Acknowledgements. ---} This work was partly conducted at
the IRT \texttt{b<>com} institute, and supported by the ANR contracts
\textsc{Articho} (Advanced Real Time Ingredients for Computational
Holography) and \textsc{Hardy} (Holographic Augmented Reality
DisplaY). The authors would like to thank Anas El Rhammad, Antonin
Gilles and Christophe Cheverry for fruitful discussions. The numerical
experiments illustrating this work have been performed using the
open-source library LTFAT~\cite{ltfatnote030}, and in particular its
Python port, \texttt{ltfatpy}~\cite{ltfatpy}.

\bibliographystyle{abbrv}%
\bibliography{bibli}
\end{document}